\documentclass[aps,prc,twocolumn,superscriptaddress, nofootinbib]{revtex4-1}[11pt]    	% revtex4-1 prc style
\pdfoutput=1
\usepackage{amsmath,amssymb,graphicx,lineno}
\usepackage{upgreek}
\usepackage{hyperref}
\usepackage{color}
\usepackage[utf8]{inputenc}
\graphicspath{ {./plots/} }

\newcommand{\kevnr}{keV$_{nr}$}

\newcommand{\scix}[3][]
{ 
 \ifthenelse{\equal{#2}{}}
  {} % True i.e. No Mantissa
  { % False i.e. There is a Mantissa
   {#2} % Mantissa
  }
 \ifthenelse{\equal{#2}{}\OR\equal{#3}{}}
  {} % True i.e. No Times required
  { % False i.e. Times required
   {\hspace{-0em} \times \hspace{-0em}} % Times
  }
 \ifthenelse{\equal{#3}{}}
  {} % No Exponent
  {10^{#3}}   % Exponent
 \ifthenelse{\equal{#1}{}}
 {}
 {\,\mathrm{#1}} % Units, optional argument
}

\begin{document}

% version 1.1 (Sept 3 2015)
% version 1.2 (Oct 5 2015)
% version 1.3 (Oct 28 2015) }
%\comment{ version 1.4 (Sept 30 2016) }
%\comment{ version 1.5 (Oct 28 2016) }
%\comment{ version 1.5.1 (Nov 2 2016) }
%\comment{ version 1.5.2 (Nov 8 2016) }
%\comment{ version 1.5.3 (Dec 8 2016) }
%\comment{ version 1.5.4 (Dec 9 2016) }
%\comment{ version 1.5.5 (Dec 8 2016) Rick editing offline }
%\comment{ version 1.5.6 (Mar 20 2017) Casey editing online }
%\comment{ version 1.5.7 (Mar 21 2017) Casey editing online }
%\comment{ ver. 1.5.8 (Mar 24 2017) DQ restructuring online }
%\comment{ ver. 1.5.10 (April 5 2017) DQ/Casey edits}
% \comment{ ver. for LUX internal review 1 (April 11 2017) }
% \comment{ ver. for LUX internal review 2 (July 7 2017) }
% \comment{ ver. for LUX collaboration review 1 (July 25 2017) }
% \comment{ ver. for LUX collaboration review 2 (August 27 2017) }
%
% 151005 Rick's initial push into GitHub and then into sharelatex.com
% Using DQ Version 1.1

\title{Ultra-Low Energy Calibration of LUX Detector using $^{127}$Xe Electron Capture}

\preprint{}

%
% Based upon the official 20170805-lux-tex-author-list.tex file:
% http://teacher.pas.rochester.edu:8080/wiki/pub/Lux/TeXAuthorList/20170805-lux-tex-author-list.tex
%

%\documentclass[superscriptaddress]{revtex4-1}
%\begin{document}

\author{D.S.~Akerib} \affiliation{Case Western Reserve University, Department of Physics, 10900 Euclid Ave, Cleveland, OH 44106, USA} \affiliation{SLAC National Accelerator Laboratory, 2575 Sand Hill Road, Menlo Park, CA 94205, USA} \affiliation{Kavli Institute for Particle Astrophysics and Cosmology, Stanford University, 452 Lomita Mall, Stanford, CA 94309, USA}
\author{S.~Alsum} \affiliation{University of Wisconsin-Madison, Department of Physics, 1150 University Ave., Madison, WI 53706, USA}  
\author{H.M.~Ara\'{u}jo} \affiliation{Imperial College London, High Energy Physics, Blackett Laboratory, London SW7 2BZ, United Kingdom}  
\author{X.~Bai} \affiliation{South Dakota School of Mines and Technology, 501 East St Joseph St., Rapid City, SD 57701, USA}  
\author{A.J.~Bailey} \affiliation{Imperial College London, High Energy Physics, Blackett Laboratory, London SW7 2BZ, United Kingdom}  
\author{J.~Balajthy} \affiliation{University of Maryland, Department of Physics, College Park, MD 20742, USA}  
\author{P.~Beltrame} \affiliation{SUPA, School of Physics and Astronomy, University of Edinburgh, Edinburgh EH9 3FD, United Kingdom}  
\author{E.P.~Bernard} \affiliation{University of California Berkeley, Department of Physics, Berkeley, CA 94720, USA} \affiliation{Yale University, Department of Physics, 217 Prospect St., New Haven, CT 06511, USA} 
\author{A.~Bernstein} \affiliation{Lawrence Livermore National Laboratory, 7000 East Ave., Livermore, CA 94551, USA}  
\author{T.P.~Biesiadzinski} \affiliation{Case Western Reserve University, Department of Physics, 10900 Euclid Ave, Cleveland, OH 44106, USA} \affiliation{SLAC National Accelerator Laboratory, 2575 Sand Hill Road, Menlo Park, CA 94205, USA} \affiliation{Kavli Institute for Particle Astrophysics and Cosmology, Stanford University, 452 Lomita Mall, Stanford, CA 94309, USA}
\author{E.M.~Boulton} \affiliation{University of California Berkeley, Department of Physics, Berkeley, CA 94720, USA} \affiliation{Yale University, Department of Physics, 217 Prospect St., New Haven, CT 06511, USA} 
\author{P.~Br\'as} \affiliation{LIP-Coimbra, Department of Physics, University of Coimbra, Rua Larga, 3004-516 Coimbra, Portugal}  
\author{D.~Byram} \affiliation{University of South Dakota, Department of Physics, 414E Clark St., Vermillion, SD 57069, USA} \affiliation{South Dakota Science and Technology Authority, Sanford Underground Research Facility, Lead, SD 57754, USA} 
\author{S.B.~Cahn} \affiliation{Yale University, Department of Physics, 217 Prospect St., New Haven, CT 06511, USA}  
\author{M.C.~Carmona-Benitez} \affiliation{Pennsylvania State University, Department of Physics, 104 Davey Lab, University Park, PA  16802-6300, USA} \affiliation{University of California Santa Barbara, Department of Physics, Santa Barbara, CA 93106, USA} 
\author{C.~Chan} \affiliation{Brown University, Department of Physics, 182 Hope St., Providence, RI 02912, USA}  
%\author{A.A.~Chiller} \affiliation{University of South Dakota, Department of Physics, 414E Clark St., Vermillion, SD 57069, USA}  
%\author{C.~Chiller} \affiliation{University of South Dakota, Department of Physics, 414E Clark St., Vermillion, SD 57069, USA}  
\author{A.~Currie} \affiliation{Imperial College London, High Energy Physics, Blackett Laboratory, London SW7 2BZ, United Kingdom}  
\author{J.E.~Cutter} \affiliation{University of California Davis, Department of Physics, One Shields Ave., Davis, CA 95616, USA}  
\author{T.J.R.~Davison} \affiliation{SUPA, School of Physics and Astronomy, University of Edinburgh, Edinburgh EH9 3FD, United Kingdom}  
\author{A.~Dobi} \affiliation{Lawrence Berkeley National Laboratory, 1 Cyclotron Rd., Berkeley, CA 94720, USA}  
%\author{J.E.Y.~Dobson} \affiliation{Department of Physics and Astronomy, University College London, Gower Street, London WC1E 6BT, United Kingdom}  
\author{E.~Druszkiewicz} \affiliation{University of Rochester, Department of Physics and Astronomy, Rochester, NY 14627, USA}  
\author{B.N.~Edwards} \affiliation{Yale University, Department of Physics, 217 Prospect St., New Haven, CT 06511, USA}  
%\author{C.H.~Faham} \affiliation{Lawrence Berkeley National Laboratory, 1 Cyclotron Rd., Berkeley, CA 94720, USA}  
\author{S.R.~Fallon} \affiliation{University at Albany, State University of New York, Department of Physics, 1400 Washington Ave., Albany, NY 12222, USA}  
\author{A.~Fan} \affiliation{SLAC National Accelerator Laboratory, 2575 Sand Hill Road, Menlo Park, CA 94205, USA} \affiliation{Kavli Institute for Particle Astrophysics and Cosmology, Stanford University, 452 Lomita Mall, Stanford, CA 94309, USA} 
\author{S.~Fiorucci} \affiliation{Lawrence Berkeley National Laboratory, 1 Cyclotron Rd., Berkeley, CA 94720, USA} \affiliation{Brown University, Department of Physics, 182 Hope St., Providence, RI 02912, USA} 
\author{R.J.~Gaitskell} \affiliation{Brown University, Department of Physics, 182 Hope St., Providence, RI 02912, USA}  
%\author{V.M.~Gehman} \affiliation{Lawrence Berkeley National Laboratory, 1 Cyclotron Rd., Berkeley, CA 94720, USA}  
\author{J.~Genovesi} \affiliation{University at Albany, State University of New York, Department of Physics, 1400 Washington Ave., Albany, NY 12222, USA}  
\author{C.~Ghag} \affiliation{Department of Physics and Astronomy, University College London, Gower Street, London WC1E 6BT, United Kingdom}  
%\author{K.R.~Gibson} \affiliation{Case Western Reserve University, Department of Physics, 10900 Euclid Ave, Cleveland, OH 44106, USA}  
\author{M.G.D.~Gilchriese} \affiliation{Lawrence Berkeley National Laboratory, 1 Cyclotron Rd., Berkeley, CA 94720, USA}  
\author{C.R.~Hall} \affiliation{University of Maryland, Department of Physics, College Park, MD 20742, USA}  
\author{M.~Hanhardt} \affiliation{South Dakota School of Mines and Technology, 501 East St Joseph St., Rapid City, SD 57701, USA} \affiliation{South Dakota Science and Technology Authority, Sanford Underground Research Facility, Lead, SD 57754, USA} 
\author{S.J.~Haselschwardt} \affiliation{University of California Santa Barbara, Department of Physics, Santa Barbara, CA 93106, USA}  
\author{S.A.~Hertel} \affiliation{University of Massachusetts, Amherst Center for Fundamental Interactions and Department of Physics, Amherst, MA 01003-9337 USA} \affiliation{Lawrence Berkeley National Laboratory, 1 Cyclotron Rd., Berkeley, CA 94720, USA} \affiliation{Yale University, Department of Physics, 217 Prospect St., New Haven, CT 06511, USA}
\author{D.P.~Hogan} \affiliation{University of California Berkeley, Department of Physics, Berkeley, CA 94720, USA}  
\author{M.~Horn} \affiliation{South Dakota Science and Technology Authority, Sanford Underground Research Facility, Lead, SD 57754, USA} \affiliation{University of California Berkeley, Department of Physics, Berkeley, CA 94720, USA} \affiliation{Yale University, Department of Physics, 217 Prospect St., New Haven, CT 06511, USA}
\author{D.Q.~Huang} \email[Corresponding Author: ]{dongqing{\_}huang@brown.edu}
\affiliation{Brown University, Department of Physics, 182 Hope St., Providence, RI 02912, USA}
\author{C.M.~Ignarra} \affiliation{SLAC National Accelerator Laboratory, 2575 Sand Hill Road, Menlo Park, CA 94205, USA} \affiliation{Kavli Institute for Particle Astrophysics and Cosmology, Stanford University, 452 Lomita Mall, Stanford, CA 94309, USA} 
\author{R.G.~Jacobsen} \affiliation{University of California Berkeley, Department of Physics, Berkeley, CA 94720, USA}  
\author{W.~Ji} \affiliation{Case Western Reserve University, Department of Physics, 10900 Euclid Ave, Cleveland, OH 44106, USA} \affiliation{SLAC National Accelerator Laboratory, 2575 Sand Hill Road, Menlo Park, CA 94205, USA} \affiliation{Kavli Institute for Particle Astrophysics and Cosmology, Stanford University, 452 Lomita Mall, Stanford, CA 94309, USA}
\author{K.~Kamdin} \affiliation{University of California Berkeley, Department of Physics, Berkeley, CA 94720, USA}  
\author{K.~Kazkaz} \affiliation{Lawrence Livermore National Laboratory, 7000 East Ave., Livermore, CA 94551, USA}  
\author{D.~Khaitan} \affiliation{University of Rochester, Department of Physics and Astronomy, Rochester, NY 14627, USA}  
\author{R.~Knoche} \affiliation{University of Maryland, Department of Physics, College Park, MD 20742, USA}  
%\author{E.V.~Korolkova} \affiliation{University of Sheffield, Department of Physics and Astronomy, Sheffield, S3 7RH, United Kingdom}  
%\author{V.A.~Kudryavtsev} \affiliation{University of Sheffield, Department of Physics and Astronomy, Sheffield, S3 7RH, United Kingdom}  
\author{N.A.~Larsen} \affiliation{Yale University, Department of Physics, 217 Prospect St., New Haven, CT 06511, USA}  
%\author{C.~Lee} \affiliation{Case Western Reserve University, Department of Physics, 10900 Euclid Ave, Cleveland, OH 44106, USA} \affiliation{SLAC National Accelerator Laboratory, 2575 Sand Hill Road, Menlo Park, CA 94205, USA} \affiliation{Kavli Institute for Particle Astrophysics and Cosmology, Stanford University, 452 Lomita Mall, Stanford, CA 94309, USA}
\author{B.G.~Lenardo} \affiliation{University of California Davis, Department of Physics, One Shields Ave., Davis, CA 95616, USA} \affiliation{Lawrence Livermore National Laboratory, 7000 East Ave., Livermore, CA 94551, USA} 
\author{K.T.~Lesko} \affiliation{Lawrence Berkeley National Laboratory, 1 Cyclotron Rd., Berkeley, CA 94720, USA}  
%\author{C.~Levy} \affiliation{University at Albany, State University of New York, Department of Physics, 1400 Washington Ave., Albany, NY 12222, USA} \affiliation{Lawrence Berkeley National Laboratory, 1 Cyclotron Rd., Berkeley, CA 94720, USA} 
%\author{J.~Liao} \affiliation{Brown University, Department of Physics, 182 Hope St., Providence, RI 02912, USA}  
\author{A.~Lindote} \affiliation{LIP-Coimbra, Department of Physics, University of Coimbra, Rua Larga, 3004-516 Coimbra, Portugal}  
\author{M.I.~Lopes} \affiliation{LIP-Coimbra, Department of Physics, University of Coimbra, Rua Larga, 3004-516 Coimbra, Portugal}  
\author{A.~Manalaysay} \affiliation{University of California Davis, Department of Physics, One Shields Ave., Davis, CA 95616, USA}  
\author{R.L.~Mannino} \affiliation{Texas A \& M University, Department of Physics, College Station, TX 77843, USA} \affiliation{University of Wisconsin-Madison, Department of Physics, 1150 University Ave., Madison, WI 53706, USA} 
%\author{N.~Marangou} \affiliation{Imperial College London, High Energy Physics, Blackett Laboratory, London SW7 2BZ, United Kingdom}  
\author{M.F.~Marzioni} \affiliation{SUPA, School of Physics and Astronomy, University of Edinburgh, Edinburgh EH9 3FD, United Kingdom}  
\author{D.N.~McKinsey} \affiliation{University of California Berkeley, Department of Physics, Berkeley, CA 94720, USA} \affiliation{Lawrence Berkeley National Laboratory, 1 Cyclotron Rd., Berkeley, CA 94720, USA} \affiliation{Yale University, Department of Physics, 217 Prospect St., New Haven, CT 06511, USA}
\author{D.-M.~Mei} \affiliation{University of South Dakota, Department of Physics, 414E Clark St., Vermillion, SD 57069, USA}  
\author{J.~Mock} \affiliation{University at Albany, State University of New York, Department of Physics, 1400 Washington Ave., Albany, NY 12222, USA}  
\author{M.~Moongweluwan} \affiliation{University of Rochester, Department of Physics and Astronomy, Rochester, NY 14627, USA}  
\author{J.A.~Morad} \affiliation{University of California Davis, Department of Physics, One Shields Ave., Davis, CA 95616, USA}  
\author{A.St.J.~Murphy} \affiliation{SUPA, School of Physics and Astronomy, University of Edinburgh, Edinburgh EH9 3FD, United Kingdom}  
\author{C.~Nehrkorn} \affiliation{University of California Santa Barbara, Department of Physics, Santa Barbara, CA 93106, USA}  
\author{H.N.~Nelson} \affiliation{University of California Santa Barbara, Department of Physics, Santa Barbara, CA 93106, USA}  
\author{F.~Neves} \affiliation{LIP-Coimbra, Department of Physics, University of Coimbra, Rua Larga, 3004-516 Coimbra, Portugal}  
\author{K.~O'Sullivan} \affiliation{University of California Berkeley, Department of Physics, Berkeley, CA 94720, USA} \affiliation{Lawrence Berkeley National Laboratory, 1 Cyclotron Rd., Berkeley, CA 94720, USA} \affiliation{Yale University, Department of Physics, 217 Prospect St., New Haven, CT 06511, USA}
\author{K.C.~Oliver-Mallory} \affiliation{University of California Berkeley, Department of Physics, Berkeley, CA 94720, USA}  
\author{K.J.~Palladino} \affiliation{University of Wisconsin-Madison, Department of Physics, 1150 University Ave., Madison, WI 53706, USA} \affiliation{SLAC National Accelerator Laboratory, 2575 Sand Hill Road, Menlo Park, CA 94205, USA} \affiliation{Kavli Institute for Particle Astrophysics and Cosmology, Stanford University, 452 Lomita Mall, Stanford, CA 94309, USA}
\author{E.K.~Pease} \affiliation{University of California Berkeley, Department of Physics, Berkeley, CA 94720, USA} \affiliation{Yale University, Department of Physics, 217 Prospect St., New Haven, CT 06511, USA} 
%\author{L.~Reichhart} \affiliation{Department of Physics and Astronomy, University College London, Gower Street, London WC1E 6BT, United Kingdom}  
\author{C.~Rhyne} \affiliation{Brown University, Department of Physics, 182 Hope St., Providence, RI 02912, USA}  
%\author{P.~Rossiter} \affiliation{University of Sheffield, Department of Physics and Astronomy, Sheffield, S3 7RH, United Kingdom}  
\author{S.~Shaw} \affiliation{University of California Santa Barbara, Department of Physics, Santa Barbara, CA 93106, USA} \affiliation{Department of Physics and Astronomy, University College London, Gower Street, London WC1E 6BT, United Kingdom} 
\author{T.A.~Shutt} \affiliation{Case Western Reserve University, Department of Physics, 10900 Euclid Ave, Cleveland, OH 44106, USA}  \affiliation{Kavli Institute for Particle Astrophysics and Cosmology, Stanford University, 452 Lomita Mall, Stanford, CA 94309, USA}
\author{C.~Silva} \affiliation{LIP-Coimbra, Department of Physics, University of Coimbra, Rua Larga, 3004-516 Coimbra, Portugal}  
\author{M.~Solmaz} \affiliation{University of California Santa Barbara, Department of Physics, Santa Barbara, CA 93106, USA}  
\author{V.N.~Solovov} \affiliation{LIP-Coimbra, Department of Physics, University of Coimbra, Rua Larga, 3004-516 Coimbra, Portugal}  
\author{P.~Sorensen} \affiliation{Lawrence Berkeley National Laboratory, 1 Cyclotron Rd., Berkeley, CA 94720, USA}  
%\author{S.~Stephenson} \affiliation{University of California Davis, Department of Physics, One Shields Ave., Davis, CA 95616, USA}  
\author{T.J.~Sumner} \affiliation{Imperial College London, High Energy Physics, Blackett Laboratory, London SW7 2BZ, United Kingdom}  
\author{M.~Szydagis} \affiliation{University at Albany, State University of New York, Department of Physics, 1400 Washington Ave., Albany, NY 12222, USA}  
\author{D.J.~Taylor} \affiliation{South Dakota Science and Technology Authority, Sanford Underground Research Facility, Lead, SD 57754, USA}  
\author{W.C.~Taylor} \affiliation{Brown University, Department of Physics, 182 Hope St., Providence, RI 02912, USA}  
\author{B.P.~Tennyson} \affiliation{Yale University, Department of Physics, 217 Prospect St., New Haven, CT 06511, USA}  
\author{P.A.~Terman} \affiliation{Texas A \& M University, Department of Physics, College Station, TX 77843, USA}  
\author{D.R.~Tiedt} \affiliation{South Dakota School of Mines and Technology, 501 East St Joseph St., Rapid City, SD 57701, USA}  
\author{W.H.~To} \affiliation{California State University Stanislaus, Department of Physics, 1 University Circle, Turlock, CA 95382, USA} \affiliation{SLAC National Accelerator Laboratory, 2575 Sand Hill Road, Menlo Park, CA 94205, USA} \affiliation{Kavli Institute for Particle Astrophysics and Cosmology, Stanford University, 452 Lomita Mall, Stanford, CA 94309, USA}
\author{M.~Tripathi} \affiliation{University of California Davis, Department of Physics, One Shields Ave., Davis, CA 95616, USA}  
\author{L.~Tvrznikova} \affiliation{University of California Berkeley, Department of Physics, Berkeley, CA 94720, USA} \affiliation{Yale University, Department of Physics, 217 Prospect St., New Haven, CT 06511, USA} 
\author{S.~Uvarov} \affiliation{University of California Davis, Department of Physics, One Shields Ave., Davis, CA 95616, USA}  
\author{V.~Velan} \affiliation{University of California Berkeley, Department of Physics, Berkeley, CA 94720, USA}  
\author{J.R.~Verbus} \affiliation{Brown University, Department of Physics, 182 Hope St., Providence, RI 02912, USA}  
\author{R.C.~Webb} \affiliation{Texas A \& M University, Department of Physics, College Station, TX 77843, USA}  
\author{J.T.~White} \affiliation{Texas A \& M University, Department of Physics, College Station, TX 77843, USA}  
\author{T.J.~Whitis} \affiliation{Case Western Reserve University, Department of Physics, 10900 Euclid Ave, Cleveland, OH 44106, USA} \affiliation{SLAC National Accelerator Laboratory, 2575 Sand Hill Road, Menlo Park, CA 94205, USA} \affiliation{Kavli Institute for Particle Astrophysics and Cosmology, Stanford University, 452 Lomita Mall, Stanford, CA 94309, USA}
\author{M.S.~Witherell} \affiliation{Lawrence Berkeley National Laboratory, 1 Cyclotron Rd., Berkeley, CA 94720, USA}  
\author{F.L.H.~Wolfs} \affiliation{University of Rochester, Department of Physics and Astronomy, Rochester, NY 14627, USA}  
%\author{D.~Woodward} \affiliation{University of Sheffield, Department of Physics and Astronomy, Sheffield, S3 7RH, United Kingdom}  
\author{J.~Xu} \affiliation{Lawrence Livermore National Laboratory, 7000 East Ave., Livermore, CA 94551, USA}  
\author{K.~Yazdani} \affiliation{Imperial College London, High Energy Physics, Blackett Laboratory, London SW7 2BZ, United Kingdom}  
\author{S.K.~Young} \affiliation{University at Albany, State University of New York, Department of Physics, 1400 Washington Ave., Albany, NY 12222, USA}  
\author{C.~Zhang} \affiliation{University of South Dakota, Department of Physics, 414E Clark St., Vermillion, SD 57069, USA} 

%\maketitle
%\end{document}

\begin{abstract}
We report an absolute calibration of the ionization yields~($\textit{Q$_y$})$ and fluctuations for electronic recoil events in liquid xenon at discrete energies between 186 eV and 33.2 keV. The average electric field applied across the liquid xenon target is 180 V/cm. The data are obtained using low energy $^{127}$Xe electron capture decay events from the 95.0-day first run from LUX (WS2013) in search of Weakly Interacting Massive Particles (WIMPs). The sequence of gamma-ray and X-ray cascades associated with $^{127}$I de-excitations produces clearly identified 2-vertex events in the LUX detector. We observe the K- (binding energy, 33.2 keV), L- (5.2 keV), M- (1.1 keV), and N- (186 eV) shell cascade events and verify that the relative ratio of observed events for each shell agrees with calculations. The N-shell cascade analysis includes single extracted electron (SE) events and represents the lowest-energy electronic recoil $\textit{in situ}$ measurements that have been explored in liquid xenon.
\end{abstract}

\maketitle

\setcounter{equation}{0} \setcounter{footnote}{0} % what this does? RG - This takes of the numbering for footnotes, if you use them

%%%%%%%%%%%%%%%%%%%%%%%%%%%%%%%%%%%%%%%%%%%%%%%%%%%
\section{Introduction}\label{sec:introduction}

The LUX dark matter search experiment~\cite{LUXexperiment} is a 250~kg active mass dual-phase (liquid/gas) xenon time projection chamber located at the 1480~m level of the Sanford Underground Research Facility in Lead, South Dakota, USA. LUX detects both scintillation and ionization signals produced by incident or internally emitted particles interacting with xenon atoms in the active region via either electronic recoil (ER) or nuclear recoil (NR).

The recoil interaction initially produces excitons (Xe$^*$) and electron-ion (e$^-$Xe${^+}$) pairs at the interaction site.
The excitons collide with neutral neighbours to form Xe$_2^*$ excited dimers which decay on a timescale of tens of nanoseconds and produce the primary scintillation light, denoted as S1. A fraction of the electrons also re-combine with ions and produce additional scintillation light contributing to S1 on a similar timescale.
The remaining electrons which survive recombination are drifted upwards by the applied vertical 
electric field in the liquid xenon (LXe) active volume. 
An electric field with a mean and range in the fiducial volume of $180\pm20$ V/cm~\cite{LUXRun3Reanalysisprl} is applied during WS2013.
The mean electron drift velocity is 
$1.51\pm0.01$~mm/$\upmu$s~\cite{LUXfirstResult}. The electrons are then extracted from the liquid to the gas phase with an extraction efficiency of 0.49$\pm$0.03~\cite{LUXprd}.
The extracted electrons subsequently undergo electroluminescence in the gas phase; this proportional scintillation light is known as S2. 
Each extracted electron induces a mean of $24.66\pm0.02$ detected photons (phd) and a 1$\sigma$ width of $5.95\pm0.02$ phd~\cite{LUXprd, LUXDD, DoublePhe} across all photomultiplier tube (PMT) photocathodes.
The event $(x, y)$ position is reconstructed from the S2 light distribution in the top PMT array~\cite{mercury}, while the $z$ position is determined based on the time delay between the S1 and S2 signals.

For NR events, in addition to excitons and electron-ion pairs, part of the deposited energy is lost to atomic motion and eventually converted into heat. The energy lost by the projectile particle to atoms in the medium is well described by the Lindhard model~\cite{Lindhard1963, Sorensen2011} down to $\sim$keV energies, and has been experimentally measured by LUX for nuclear recoils in LXe over the range $0.7-74$ \kevnr ~\cite{LUXDD}. 
The ER and NR events are typically discriminated by the logarithmic charge to light ratio, i.e. log$_{10}$(S2/S1), thanks to the different ionization/excitation ratios for ER and NR interactions~\cite{Aprile2006a,MatthewNEST}.
We expect WIMPs to interact with LXe via nuclear recoil, depositing up to $\mathcal{O}(100)$~keV in a single scatter. 
LUX has reported world-leading dark matter search results on both spin-independent and spin-dependent WIMP-nucleon scattering in~\cite{LUXfirstResult, LUXRun3Reanalysisprl, spinDep, run4prl}.

%The result of Run3, that was carried out between April and August of 2013, has been reported in~\cite{LUXRun3Reanalysisprl}, in which LUX is demonstrated to be the world's most sensitive dark matter search experiment to date by setting most stringent WIMP-nucleon elastic cross section $90{\%}$ confident limit with a minimum of  0.6~zb at a WIMP mass of $33$~GeV/c$^{2}$.

In the context of a WIMP search experiment using a LXe target, it is important to understand LXe scintillation and ionization yield responses over the WIMP search energy range for both ER and NR because of their non-linear energy dependence~\cite{Doke2002, KaiX2006}. Many efforts have been devoted to understanding the scintillation and ionization response in LXe in the past few years using various techniques~\cite{Sorensen2009339, Manzur, Horn:2011wz, Lim, Baudis}. LUX has independently developed and deployed a number of novel $\textit{in situ}$ internal and external sources to calibrate detector ER and NR response in the energy region that is relevant to WIMP searches. Two such sources are tritiated methane (CH$_{3}$T) for ER calibration~\cite{LUXTritium} and deuterium-deuterium (D-D) neutrons for NR calibration~\cite{LUXDD}. While tritium is an ideal source to calibrate detector ER response in the low energy region, its application is limited by it being a continuum-energy source which affects the sensitivities at low energies, and the detector light collection efficiency. As a result, the tritium calibration currently reaches a lowest-energy calibration point of 1.3~keV~\cite{LUXTritium}. A source that is capable of studying calibrations in the sub-keV energy range in LXe is desirable. For example, this small signal regime is directly relevant to the signal and backgrounds for low-mass WIMP searches and for coherent neutrino-nucleus scattering (CNNS)~\cite{CNNS, Baudis:2013qla}.

\section{Xenon-127 in LUX Detector}\label{sec:xe127}
LUX background measurements with WS2013 data revealed an initial $^{127}$Xe activity of 
$490\pm95$ ${\upmu}$Bq/kg 
in the active region~\cite{LUXBG}. From this, we infer approximately 0.8~million $^{127}$Xe decay events during the WS2013 3-month run period, given the 36.4~day half-life of the isotope. 
The $^{127}$Xe radioisotope is present in the LXe target due to cosmogenic activation of the Xe during its time on the surface 
before being brought one mile underground. 
The surface production rate is modeled and estimated using ACTIVIA and described in~\cite{LUXBG}. 
The decay characteristics of $^{127}$Xe make it an appealing mono-energetic source for LUX ER energy calibration. This calibration covers the entire signal region relevant to the WIMP search, reaching all the way down to the observation of 186~eV energy deposition. This represents the lowest-energy ER $\textit{in situ}$ measurements that have been explored in LXe to date. 

\begin{figure}[htbp]
\begin{center}
\vskip -0.3cm % -0.5
\includegraphics[width=0.45\textwidth]{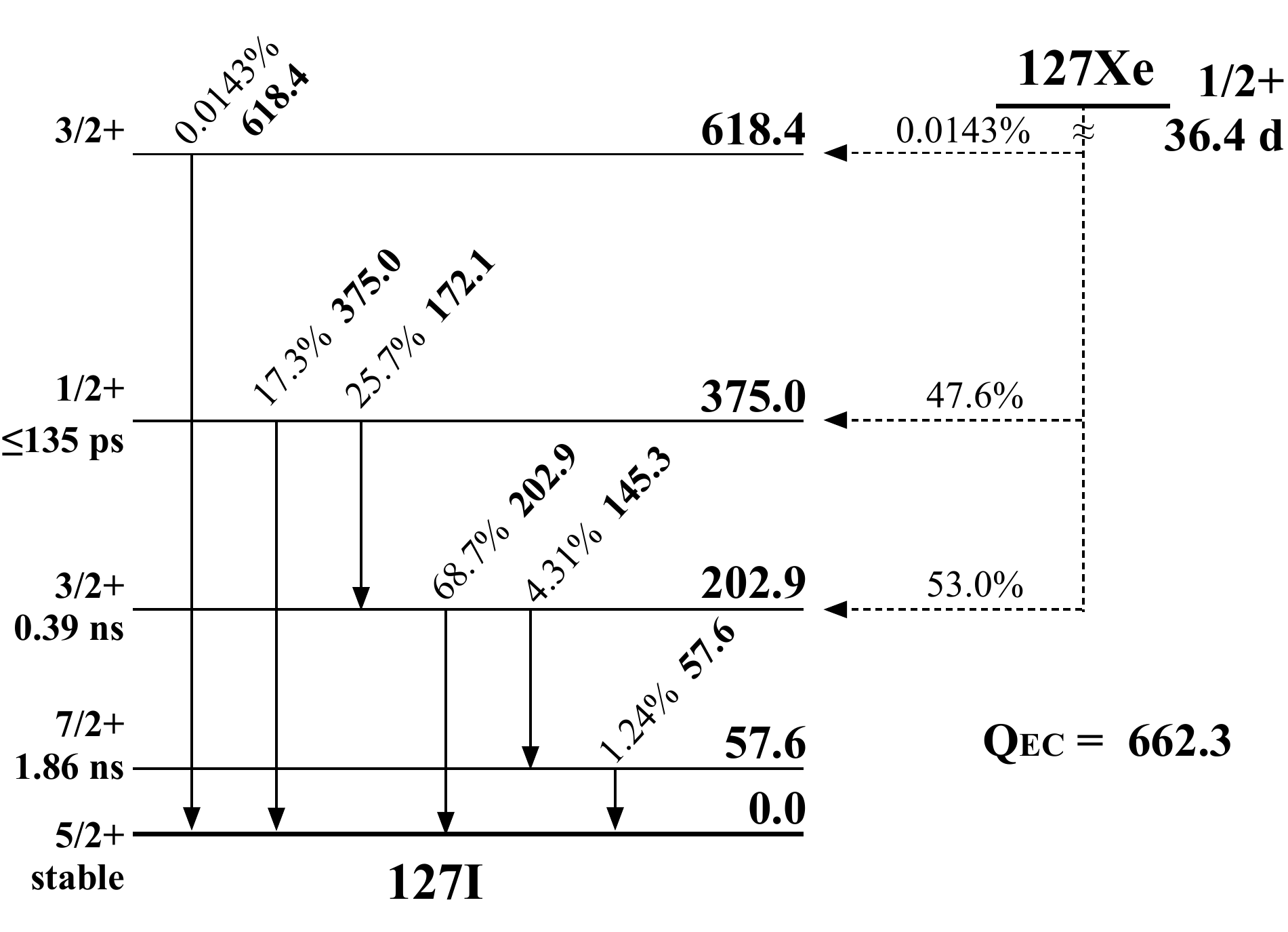} %0.42
\vskip -0.1cm
\caption{Decay scheme of $^{127}$Xe~\cite{127xedecay} with units of keV. The $^{127}$Xe decays via electron capture to $^{127}$I. The percentage above the transition arrow is the gamma-ray intensity as fraction of parent ($^{127}$Xe) decay.
}
\vskip -0.5cm
\label{fig:decay_scheme_127Xe}
\end{center}
\end{figure}

The $^{127}$Xe decays via electron capture (EC), in which its nucleus absorbs one of the atomic electrons. Following this EC, the possible initial states and subsequent decays of the daughter nucleus, $^{127}$I, are shown in Fig.~\ref{fig:decay_scheme_127Xe}. The $^{127}$I is left in its 375 keV or 203 keV excited state with 47{\%} and 53{\%} probability, respectively. There is a 17.3{\%} probability of decay from the 375~keV state to ground state by a single gamma-ray emission and a 43.9{\%}~\cite{127xedecay} probability of decay from the 203~keV state to ground state via a single gamma-ray emission. 
Nuclear de-excitation can also occur via internal conversion (IC) electron emission; however, this process occurs with a branching ratio of less than 10{\%} relative to the gamma-ray emission~\cite{xray_ratio}.

\begin{figure}[htbp]
\begin{center}
\vskip -0.3cm % -0.5
\includegraphics[width=0.45\textwidth]{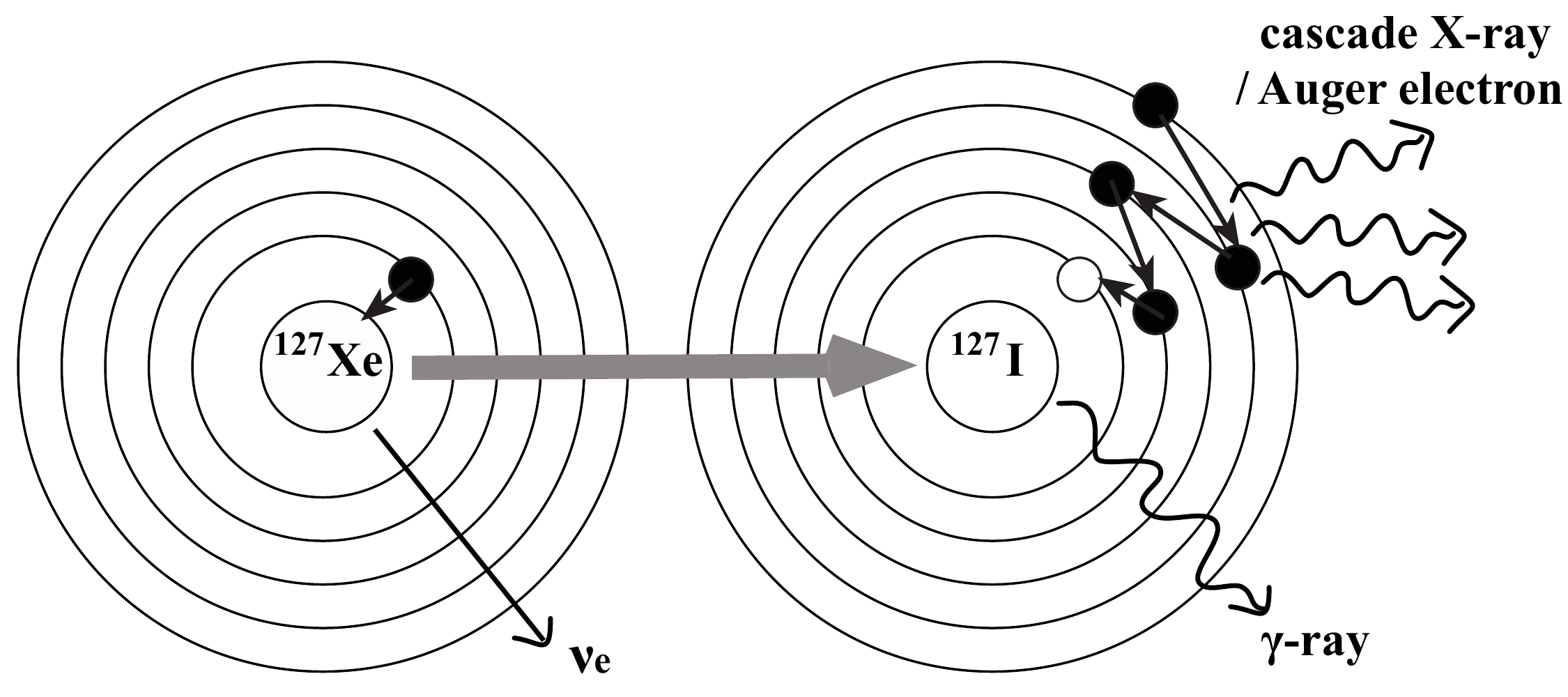}
\vskip -0.1cm
\caption{
% Toy model of $^{127}$Xe EC decay. 
A schematic illustrating atomic electron capture (a K-shell electron in this case),
for a $^{127}$Xe nucleus, which
 is converted into $^{127}$I in an excited state. 
 The excited $^{127}$I nucleus can subsequently de-excite via emission of one or more gamma rays (or IC electrons). The atomic structure de-excites through X-ray (or Auger electron) cascade emissions.
 % The nuclear de-excitation can also be through internal conversion with high-speed electron emission, but this is not relevant for this work due to short mean free path.
 }
\vskip -0.5cm
\label{fig:decay_toy_model_127Xe}
\end{center}
\end{figure}

The electron capture can occur from either the K, L, M, or N shell with 83.37{\%}, 13.09{\%}, 2.88{\%} and 0.66{\%} probabilities (see Table~\ref{tab:ratio}), respectively, resulting in an atomic orbital vacancy~\cite{xray_ratio}. 
The vacancy is subsequently filled with an electron 
from a higher level via emission of cascade X-rays or Auger electrons (Fig.~\ref{fig:decay_toy_model_127Xe}), with total cascade energies of 32.2~keV, 5.2~keV, 1.1~keV, and 186~eV~\cite{xray_energy}, respectively. Localized energy depositions associated with these processes are clearly observed by the LUX detector and are used for low and ultra-low energy ER calibration.

% Nuclear de-excitation can also take place via internal conversion (IC) electron emission or multiple gamma emission.

Our analysis focuses on the $^{127}$Xe decay events that involve a single gamma-ray emission followed by an atomic cascade. The two energy depositions are sufficiently spatially separated to be individually identified in the LUX detector. 
The IC electrons are not considered in this work due to their short range in Xe, which causes the nuclear and subsequent atomic de-excitation signals to always spatially overlap~\cite{mercury}.
The sub-dominant component of decays with multiple gamma-ray emission are not considered, as the complexity of their event energy reconstruction leads to unnecessary systematic uncertainties in the analysis. 

The nuclear and atomic de-excitations of $^{127}$I %following the $^{127}$Xe EC through respective gamma and X-ray emissions 
can be treated as prompt (ns timescale~\cite{127xedecay}) and simultaneous processes in the LUX detector, given the 
subsequent Xe scintillation light (S1) emission with timescales characterized by 10's of ns
and the 
data acquisition (DAQ) system's sampling interval of 10~ns~\cite{LUXDAQ}. 
The simultaneity is confirmed by data which shows that the ER primary S1 signals from both processes overlap with each other in time. Therefore, for a given EC event, 
%as long as the gamma does not escape the active volume, which is true for the vast majority~($>$80${\%}$) of them, 
there are two simultaneous ER processes in the active volume: one due to the gamma ray and the other due to the X-ray. Events of this type are known as double-scatter (or two-vertex) events, distinguished from single-scatter events in which there is only one particle interacting with LXe once, such as WIMP-Xe interaction. 
% Events that have two or more scatters are known as multiple-scatter events.
%Other examples of multiple-scatter events include an incident high energy gamma depositing energy multiple times via Compton scattering in LXe or a D-D neutron scattering off multiple Xe nuclei before leaving the active volume~\cite{LUXDD}. 
% *** DQ 170322 - command out the following section
%LUX analysis has been optimized to ensure it can efficiently detect multi-scatter events due to a well-configured DAQ system. It digitizes and records the waveform in each channel that rises above the detection threshold. The threshold is chosen to acquire 95${\%}$ of all single photoelectrons while suppressing baseline~\cite{LUXDAQ}. The per-channel-based digitization is independent of the trigger system; instead, the LUX trigger system~\cite{LUXtrigger} utilizes digitized pulses from all 122 PMT channels to makes trigger decisions and flags potential WIMP or source events. As the data stream is mostly occupied by baseline, this technique saves a significant amount of data volume and reduces the DAQ dead time. Each digitized pulse is known as a POD, i.e. pulse only digitization. The S1 and S2 pulse waveforms are then reconstructed via summing up PODs across all PMT channels. 
% For more details, see the referred papers. 
%Most importantly, the POD data acquisition mode preserves a complete set of information for all events, including $^{127}$Xe EC double scatters.

\begin{figure}[htbp]
\begin{center}
\vskip -0.3cm % -0.5
\includegraphics[width=0.45\textwidth]{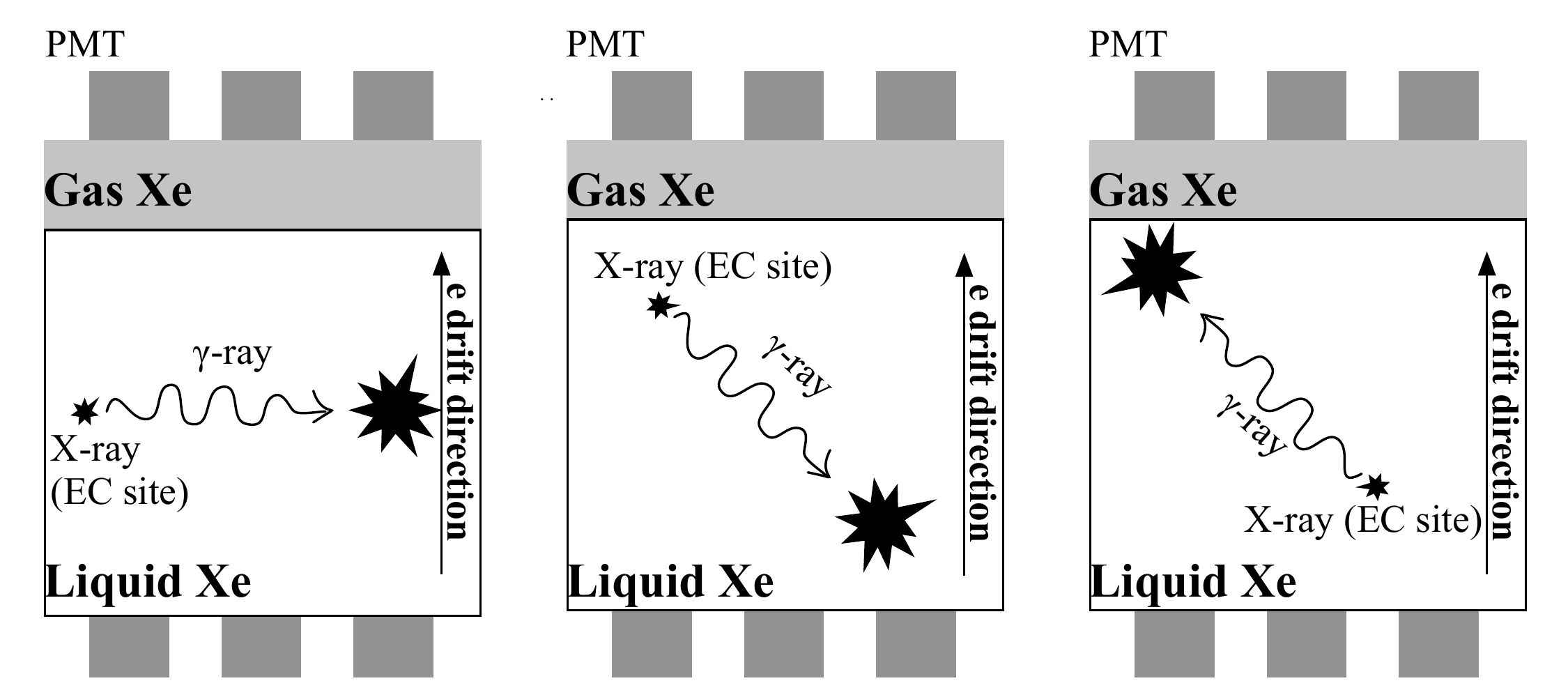}
\vskip -0.1cm
\caption{Schematics (not to scale) of $^{127}$Xe decay events in the LUX detector where both the X-ray and gamma ray have ER interactions in the active volume. 
Due to the relatively short MFP, the X-ray ER interaction site is considered the same as where the initial nuclear EC happens.
Depending on the component of the gamma-ray travel in the vertical direction, the subsequent drift-readout of the event in the S2 can appear as two S2s merged with each other (left), as a small (X-ray deposition) S2 followed by a large (gamma-ray deposition) S2 (middle), or a large S2 followed by a small S2 (right).}
\vskip -0.5cm
\label{fig:event_schematic_127Xe}
\end{center}
\end{figure}

The mean free path (MFP) in the LXe for gamma rays at 203~keV and 375~keV 
is 0.93~cm and 2.56~cm~\cite{photonAsorpLength}, respectively.
The EC X-ray, which has the maximum possible energy of 32.2~keV, has a MFP of ${<}$ 0.05~cm~\cite{photonAsorpLength} in LXe. 
In this analysis, the X-ray ER interaction site can be considered to be at the same location as where the initial nuclear EC occurs. 
The relative spatial location of two ER interactions sites are therefore predominantly defined by the gamma-ray travel direction and distance in the LXe volume. 
Schematics of a typical $^{127}$Xe EC event in the LUX detector are shown in Fig.~\ref{fig:event_schematic_127Xe}.

\begin{figure}[htbp]
\begin{center}
\vskip -0.3cm % -0.5
\includegraphics[width=0.45\textwidth]{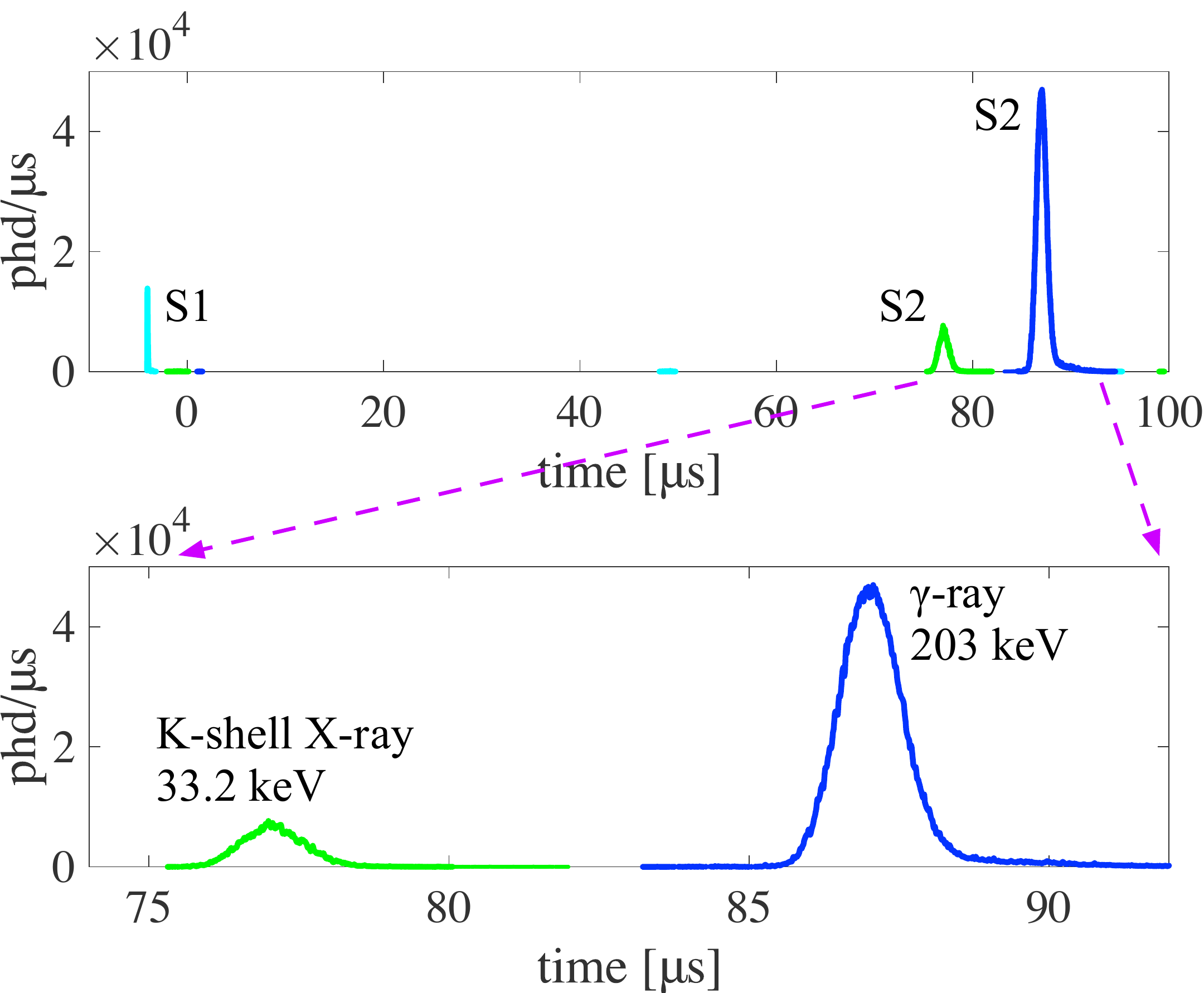}
\vskip -0.1cm
\caption{The time trace, using 10 ns samples, in LUX summed across all 122 PMTs for a $^{127}$Xe decay event with K-shell electron capture in LUX data (top).
The event appears as a clear double-scatter.
An S1 signal is followed by two well separated S2s. 
The first (green) S2 is due to K-shell X-ray ER, and the second (blue) S2 is due to the 203~keV gamma-ray ER in the LXe volume. 
The two S1s from both ERs overlap and appear as one S1 (cyan). 
The (bottom) plot is a closeup of the two S2s. 
%The unit, phe, on the $y$-axis stands for photoelectron, the unit used before signal is converted to detected photon (phd)
}
\vskip -0.5cm
\label{fig:real_xe127_event}
\end{center}
\end{figure}

The X-ray and gamma ray independently produce both scintillation (S1) and ionization (S2) signals at their ER interaction sites. 
The two S1 signals originating from these sites cannot be separately resolved in LUX data (Fig.~\ref{fig:real_xe127_event}) as discussed above. As a result, low energy ER scintillation yield ($\textit{L$_y$}$) measurements using EC double-scatter events are not possible.
Both charge signals are drifted vertically upwards to the liquid surface, and are then 
both
extracted into the gas phase to produce S2 signals. 
Depending on the relative depths in the LXe target
of the X-ray and gamma-ray ER components 
% i.e. X-ray and gamma respective ER interaction $z$ positions in LXe Volume
(Fig.~\ref{fig:event_schematic_127Xe}), 
the two S2 signals can be either well separated in drift time (reflecting their separation in depth, the $z$-coordinate)
or sometimes merged into one pulse in the reconstructed event waveform.
%In this analysis no attempt is made to use the $(x, y)$ separation of the events, since the effective $z$ position resolution is far superior. 
In the case of two S2s overlapping in drift time, the double-scatter event will be classified (in the LUX data processing framework) as a single-scatter event with a single S2 pulse, making it difficult to extract the X-ray signal.
%Since the LUX event classification algorithm is only based on the number of S1 and S2 in an event. 
% DQ 170322 - command out the following section
%The gamma S2 size can be significantly greater in magnitude than that of the X-ray S2, giving it a stronger tendency to cover up the subsequent X-ray.
Consequently, events with well separated S2 pulses along the $z$-axis are desirable, especially for the two lowest-energy X-ray calibration points, to achieve a $\textit{Q$_y$}$ measurement with minimum systematic uncertainty. 
A typical LUX $^{127}$Xe decay event via K-shell EC with two ER sites well separated in $z$ direction is presented in Fig.~\ref{fig:real_xe127_event}.
The $z$ separation is 10.0~$\upmu$s in drift time, corresponding to a distance of 15.1~mm.
% *** DQ 170322 - command out the following section
%Events with two EC vertices at the same $z$ position but well separated %in $(x, y)$ could also potentially be analyzed to extract individual S2 %pulse size. However, this fitting of the event light hit map across the %top PMT array was not implemented in this analysis, and would be %estimated to make a very modest difference to useful event statistics, %since the effective $z$ position resolution is far superior to the $(x, %y)$ resolution.
%

% ***DQ - RG 161108 Can you please alter my new text here - the earlier text didn't appear quite right - the S2 signal events aren't as wide as 3-5 us - the DAQ events may be?
% ***RG - DQ 161115 - added 1.4 us on average
%{ *** DQ 170322 - command out the following section
%The S2 pulse is contained within a Gaussian-like time envelope with a %10-90\% width of 1.4~${\upmu}$s on average. %**** What is the typical range here?
% Typical S2 pulses in LUX have average span of 3 - 5~${\upmu}$s in the $z$ direction with slight energy dependence.
%The total event span can extend to 10~${\upmu}$s due to a small amount %of additional signal that can appear in the pulse tails, especially for higher energy pulses (e.g. EC gammas). 
%A drift time of 1~${\upmu}$s is equivalent to 1.51~mm given the mean electron drift velocity of 1.51~mm/${\upmu}$s. 
%In our analysis a vertical separation of $\geq$~0.9~cm (6~${\upmu}$s) was required between two vertices to define EC events with clearly separated S2 pulses.
%} 
As the relative spatial location of the two vertices is predominantly determined by the gamma-ray propagation distance, the efficiency for observing events involving the 375~keV gamma ray
is greater because its MFP in LXe is a factor of 2.8 longer than that of the 203~keV gamma ray. However, the 375~keV gamma ray has a Compton scattering cross section a factor of 1.8~\cite{photonAsorpLength} greater than that of photoelectric absorption in LXe, which results in more triple- (or higher-multiple) scatter events and significantly reduces the double-scatter event rate. 
The 203~keV gamma-ray interactions with LXe are dominated by photoelectric absorption via full energy deposition. In addition, the branching ratio for decay with a single 203~keV gamma-ray emission is a factor 2.5 higher than that of a single 375~keV gamma-ray emission (Fig.~\ref{fig:decay_scheme_127Xe}). 
Given these considerations, EC events tagged by the 203~keV gamma-ray emission were chosen for ER calibration due to an expected higher event rate. 
About 15${\%}$ of all EC events with the 203~keV gamma-ray emission are expected to have two clearly separated S2 pulses suitable for our analysis. A total of 0.8 million EC events during WS2013 provides sufficient data for clear observation of K-, L-, M- and N-shell events for energy calibration.

\section{analysis and results}\label{sec:analysis_and_results}
\subsection{K-, L-, M- and N-Shell X-ray Analysis}\label{sec:klm}
As discussed in the previous section, 
an ideal EC event for low energy calibration contains two scatters, with one S1 followed by two well separated S2s. The size of one S2 is expected to be significantly greater than that of the other (Fig.~\ref{fig:real_xe127_event}). Depending on the gamma-ray travel direction relative to the de-excited nucleus (Fig.~\ref{fig:event_schematic_127Xe}), it can either create a large gamma-ray S2 followed by a small X-ray S2 or vice versa. The isotropic distribution of the gamma-ray emission direction makes these two scenarios equally likely to occur. 
%This symmetry can be seen in the data shown in the scatter plot of Fig.~\ref{fig:frc_f1}, where the area of the first time-ordered S2 is plotted versus the area of the second S2 for all double scatter events selected from the cuts discussed in the next paragraph.

\begin{figure}[htbp]
\begin{center}
\vskip -0.3cm % -0.5
\includegraphics[width=0.45\textwidth]{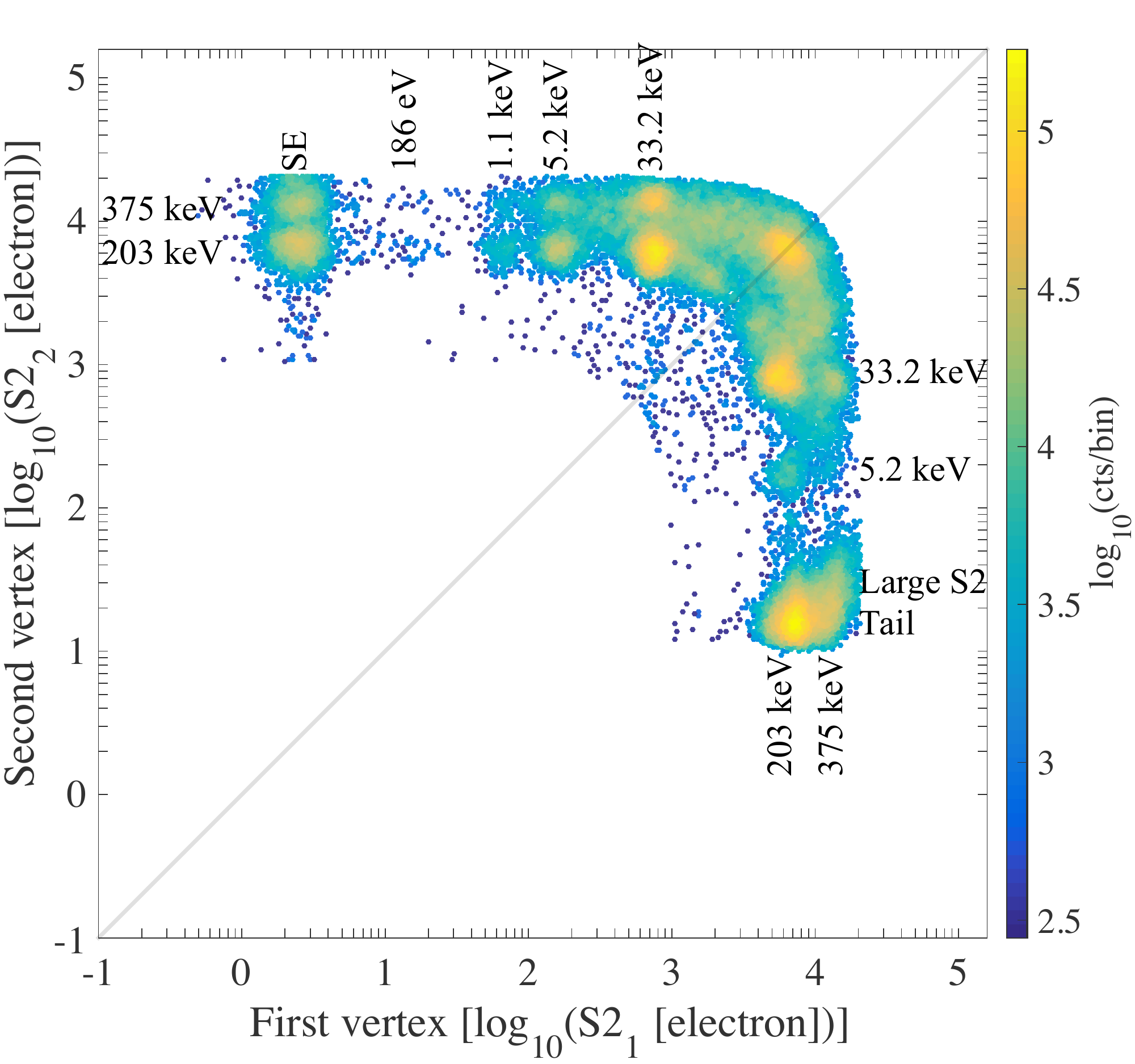}
\vskip -0.1cm
\caption{Double-scatter events associated with $^{127}$Xe decays. The ``First vertex'' is the first S2 ordered by drift time. 
The S2 sizes have been converted to the number of electrons initially drifted away from the interaction site. 
% The symmetry of the events with respect to the line $y = x$ (gray solid line) is due to isotropic distribution of gamma emission direction.  
The two arced loci of events with higher and lower signal sizes are due to the de-excitation from 375~keV-state and 203~keV-state, respectively. 
The first four populations to the left of the line 
S$2_1=~$S$2_2$ 
on the 203~keV and 375~keV bands are EC double scatter events with well isolated K-, L-, M-, and N-shell X-ray S2s from right to left. 
% These events are the main focus of this work. 
The populations labelled SE represent a background of single extracted electrons that are emitted from the liquid surface into the gas.
This later feature is not centered on log$_{10}$(S2$_1$~[electron])~=~0 because efficiency corrections are made to calibrate the corresponding signals at the interaction sites (see Sec.\ref{sec:n} for more details).
% On average this is equivalent to   is not exactly lined up with 0 on the $x$-axis (one electron) due to the electron lifetime correction applied on them for the purpose of this work.
}
\vskip -0.5cm
\label{fig:frc_f1}
\end{center}
\end{figure}

The WS2013 dataset is used~\cite{LUXRun3Reanalysisprl} for this analysis. Weekly $^{83m}$Kr calibration is performed to determine free electron lifetime~\cite{LUXKr}. The nearest $^{83m}$Kr calibration is used for S2 size correction.
Events containing exactly one S1 and two S2s are selected. An S2 threshold of 250~phd (corresponding to 10 electrons extracted from the liquid surface in the TPC) is set for the selection of the gamma-ray S2s. No pulse size threshold is applied to the selection of the X-ray S2s.
%*** DQ - 161108 RG We need to put in a reference to explain phd?
%*** RG - 161115 DQ reference added; at the first time phd appear
% The 250~phd (raw) threshold removes single-extracted-electron (SE) pulses 
%*** DQ - 161108 RG If this is true then why does we see SE in Fig 5?? Need to alter wording I think? - We need to find a different language than sub threshold because it implies that they can't be detected. Should we have Threshold A and  Threshold B?
%and small tail pulses following a large S2 (e.g. gamma) and thus increases the finding efficiency of genuine EC double-scatter events. 
%This threshold is more than 3$\sigma$ below the mean of the S2 size of the 1.1 keV M-shell X-ray (demonstrated in later text) and, thus, it has no impact on the efficiency of detection of K-, L-, and M-shell X-ray S2 distributions. 
%It will also be shown later (Sec.~\ref{sec:n}) that this is also a perfect boundary
%*** DQ - 161108 RG What does perfect boundary mean? Text below is very confusing. We need to work on it.
%to set for searching N-shell (186~eV) X-ray by implementing a separate technique. 
%In this work, any S2-like pulses with a raw size less than the set S2 threshold are labeled sub-threshold S2 pulses. 
%Sub-threshold S2 pulses includes SE pulses. 
%*** RG - 161108 RG I need to come back to this text below and work on it
% Events with $(x, y)$ and $z$ separation between two vertices within 6~cm and 4.5~cm, respectively, are considered.
Events with separation between two vertices less than 6~cm in $(x,y)$ and less than 30~${\upmu}$s (equivalent to 4.5~cm) in $z$ are considered.
This cut has 99${\%}$ acceptance for EC events that occur from a 203~keV gamma-ray emission, given a gamma-ray MFP of 0.93~cm. Because of the distinct signature of $^{127}$Xe EC decay, events can be identified with a negligible amount of background contamination. The radial fiducial cut is placed at 22~cm in this work (20~cm for~\cite{LUXRun3Reanalysisprl}), 2 cm away from the lateral detector surface to prevent potential signal charge loss to the wall. The vertical fiducial cut is kept the same as in~\cite{LUXRun3Reanalysisprl}, i.e. between 38 and 305 ${\upmu}$s in charge drift time. 
The total fiducial mass used is 178 kg, 21${\%}$ more than in~\cite{LUXRun3Reanalysisprl}. 
The double-scatter event position used for the application of the fiducial cut is defined as the energy-weighted average position of both vertices. An event total energy cut based on S2 sizes is applied to make sure that selected events are in the energy region of interest.

All double-scatter events after applying cuts are displayed in the scatter plot of Fig.~\ref{fig:frc_f1}, where the area of the first time-ordered S2 is plotted versus the area of the second S2.
% *** DQ 170323 - command out the following section
% The plot also contains events with sub-threshold S2 pulses
%*** RG - 161108 RG We need to find a different language than sub threshold because it implies that they can't be detected. Should we have Threshold A and  Threshold B??
%followed by a gamma S2 or a combined X-ray and gamma S2 (upper left populations, i.e. 186~eV and SE populations) which are added separately for the sake of completeness. 
%The detail of this will be discussed in Sec.~\ref{sec:n} for N-shell X-ray analysis.
Events to the left of the line 
S$2_1=~$S$2_2$ 
have a small S2 followed by a large S2 (Fig.~\ref{fig:event_schematic_127Xe} middle), while events to the lower-right have the opposite drift time ordering (Fig.~\ref{fig:event_schematic_127Xe} right). 
The symmetry feature with respect to the line S$2_1=~$S$2_2$ in Fig.~\ref{fig:frc_f1} is due to isotropic distribution of gamma-ray emission direction as discussed above. 
The absence of well-resolved EC peaks from the M and N shells to the lower-right of this line 
% S$2_1=~$S$2_2$ 
is caused by the extended tails (in time) of large S2 pulses, which tend to overlap with the subsequent small S2 pulses.
% This reduces the efficiency for observing double-scatter events for this polarity of event.
This reduces the efficiency for observing double-scatter events when gamma-ray S2 precedes the X-ray S2 in time.
Events with small S2s ahead of large S2s 
 (shown in Fig.~\ref{fig:event_schematic_127Xe} middle and in Fig.~\ref{fig:frc_f1} as events to the left of the line S$2_1=~$S$2_2$), 
can be more cleanly identified with well 
characterized efficiencies and so we focus on them in the rest of the EC X-ray peak analysis.
%*** DQ - 161108 RG This next section appears to come later than makes sense given discussion above - we need to rearrange order a little

The two arced bands in Fig.~\ref{fig:frc_f1} are denoted as the 203~keV band and the 375~keV band. 
%The 375~keV band contains EC events from nuclear de-excitation from the 375~keV $^{127}$I-state to the ground state, while the 203~keV band contains EC events de-exciting from the 203~keV $^{127}$I-state.
On the 203~keV band to the upper-left of the line 
S$2_1=~$S$2_2$, there are four distinct populations from right to left, which are EC double-scatter events with well isolated K-, L-, M-, and N-shell X-ray S2s, respectively, which are target events of this work. 
The four X-rays on the 375~keV band are less well-resolved due to lower statistics for reasons discussed in the previous section. Other features seen in Fig.~\ref{fig:frc_f1} are discussed in Sec.~\ref{sec:systematics}.
The EC events permit the measurement of the gamma-ray MFP and the relative ratio of observed events for each shell.
\begin{figure}[htbp]
\begin{center}
\vskip -0.3cm % -0.5
\includegraphics[width=0.45\textwidth]{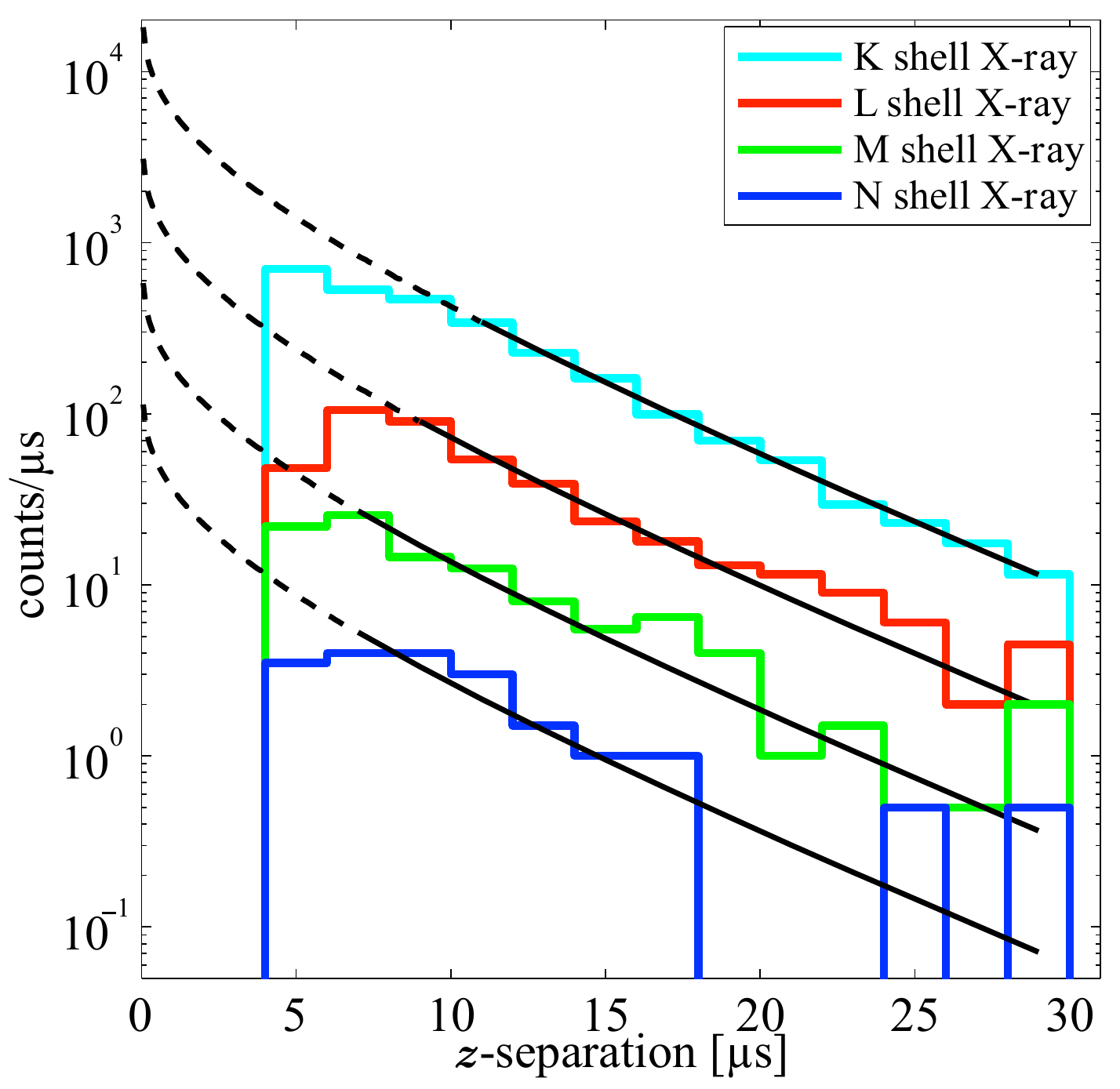}
\vskip -0.1cm
\caption{Histograms of vertical separation (1~${\upmu}$s is equivalent to 1.51~mm) between two vertices of EC events with the 203~keV gamma-ray emission for K- (cyan), L- (red), M- (green), and N- (blue) shell X-rays. The distributions are fitted with the model described by Eq. \ref{eq:semi-exp}. The solid black curves are the fits to data. The dashed black curves are the extrapolations from those fits.}
\vskip -0.5cm
\label{fig:frc_f6}
\end{center}
\end{figure}
Histograms of the vertical $z$ separation between the vertices of EC events with the 203~keV gamma-ray emission for K-, L-, M-, and N-shell X-rays are displayed in Fig.~\ref{fig:frc_f6}. Only vertical separation is used because LUX has far superior $z$ position resolution compared to $(x, y)$. In order to minimize cross-contamination from 375~keV band events due to their 2$\times$ longer gamma-ray MFP, only events with S2 sizes below the 203~keV-band mean are selected in this MFP measurement. Since the histogram represents the vertical separation of two vertices rather than the exact spatial distance, the fitting model is not exactly exponential. Instead, it is the projection of a 3-D exponential function onto one direction, as described by Eq.~\ref{eq:semi-exp}
\begin{equation} \label{eq:semi-exp}
%p (z) = \int_x\int_yexp^ (-\frac{\sqrt{ (x^2+y^2+z^2)}}{\lambda}),
   p (z) = \int_y\int_xexp{ (-\frac{x^2+y^2+z^2}{\lambda})dxdy},
\end{equation}
where $x$, $y$, and $z$ are the spatial coordinates of the 203~keV gamma-ray ER site with respect to the EC site, and $\lambda$ is the MFP of the 203~keV gamma ray. The 203~keV gamma-ray MFP is measured by fitting the K-shell distribution with a model based on Eq.~\ref{eq:semi-exp} using the least squares method. Bins which have $>$99${\%}$ efficiency, with minimum event loss due to X-ray and gamma-ray signal merging effects, are used for fitting. The best fit gives a 203~keV gamma-ray MFP of $1.04\pm0.03\pm0.10$~cm by assuming a constant 1.51~mm/${\upmu}$s electron drift velocity. The value 0.03~cm is the statistical error measured from fitting, while 0.10~cm is the systematic error due to the drift time resolution and uncertainties from LXe density, drift velocity, and X-ray location. The value is consistent with the expected value for the MFP of 0.93~cm. 
The L-, M-, and N-shell histograms are fitted with the same curve shape obtained from the K-shell fit. This is justifiable, given the K-shell X-ray MFP in LXe is 0.5~mm (0.33~${\upmu}$s electron vertical drift when X-ray travels vertically), while the S2 pulse has a 10-90$\%$ width of 1.4~${\upmu}$s on average.
\begin{table}[h]
\caption{
The observed intensities of K-, L-, M-, and N-shell EC X-rays as fraction of parent ($^{127}$Xe) decays. ``Events'' is the number of events from each shell plotted in histograms in Fig.~\ref{fig:frc_f6}; ``Amplitude'' is the $y$-intercept of each fit, which is proportional to the total number of events under each curve. The quoted errors on ``Amplitude'' are the statistical errors based on the number of events for each fit. The measured relative intensity for each shell is compared with the expected rate~\cite{xray_ratio}.}
\begin{tabular}{l@{ }r|r|r@{$\pm$}r|r|r@{$\pm$}l}
\hline \hline
%	& K 33.2~keV & L 5.2~keV & M 1.1~keV & N 186~eV \\ \hline
%Events & 2067 & 542 & 164 & 31\\ \hline
%Amplitude & $18189\pm400$ & $3089\pm133$ & $580\pm45$ & $133\pm23$ \\ \hline
%Expected ($\%$) & 83.37 & 13.09 & 2.88 & 0.66\\ \hline
%Observed ($\%$) & 82.71$\pm$2.42 & 14.05$\pm$0.66 & 2.64$\pm$0.21 & 0.60$\pm$0.11 \\
\multicolumn{2}{c|}{}	& Events & \multicolumn{2}{c|}{Amplitude} & Expected ($\%$) & \multicolumn{2}{c}{Observed ($\%$)} \\ \hline
K & 33.2~keV &  2067 & 18200 ~&~400 & 83.37 & 82.7 ~&~2.4\\ 
L & 5.2~keV  &   542 &  3090 ~&~130 & 13.09 & 14.1 ~&~0.7 \\ 
M & 1.1~keV  &   164 &   580 ~&~50  &  2.88 &  2.6 ~&~0.2\\ 
N & 186~~eV  &    31 &   133 ~&~23  &  0.66 &  0.6 ~&~0.1 \\
\hline \hline
\end{tabular}
\label{tab:ratio}
\end{table}

It is apparent in Fig.~\ref{fig:frc_f6} that a majority of EC events are missing at low $z$-separation. This is due to X-ray and gamma-ray signals overlapping. This effect is energy dependent. The underlying total number of EC events for each shell can be extrapolated. The area under the curve represents the total number of EC events, which is linearly proportional to the amplitude given by the fits of the same curve shape. The relative event ratio for each shell is estimated using the amplitudes and has good agreement with the expected rate percentages~\cite{xray_ratio}. The details are shown in Table~\ref{tab:ratio}.

\begin{figure}[htbp]
\begin{center}
\vskip 0.2cm % -0.5
\includegraphics[width=0.45\textwidth]{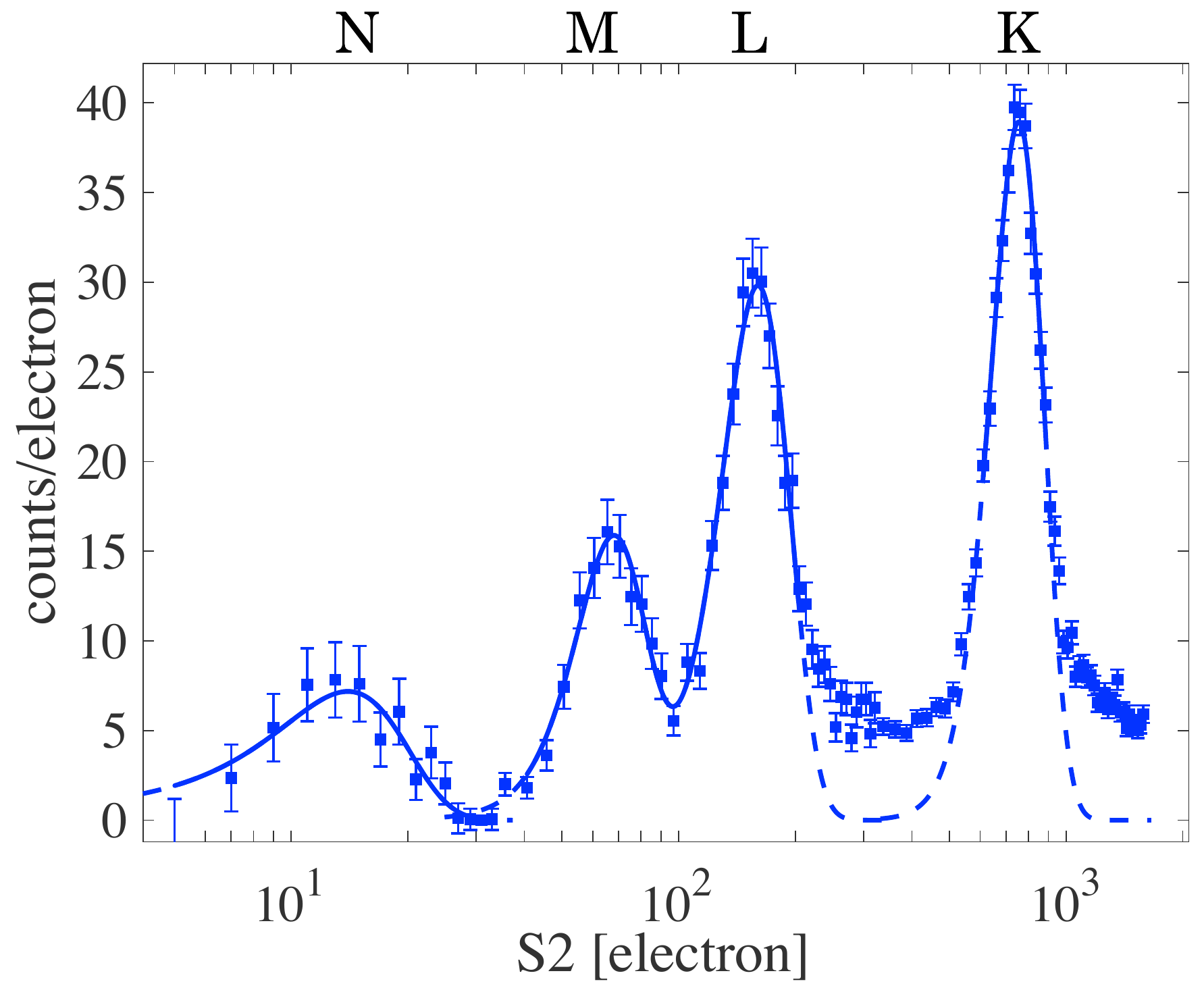}
\vskip -0.1cm
\caption{X-rays' ER charge spectra. The S2 size has been converted to the number of electrons escaping recombination at the interaction site. The peaks from right to left are due to K-, L-, M-, and N-shell X-rays respectively. The fits shown use Gaussian functions.}
\vskip -0.5cm
\label{fig:frc_f5}
\end{center}
\end{figure}

\begin{figure}[htbp]
\begin{center}
\vskip 0.1cm % -0.5
\includegraphics[width=0.45\textwidth]{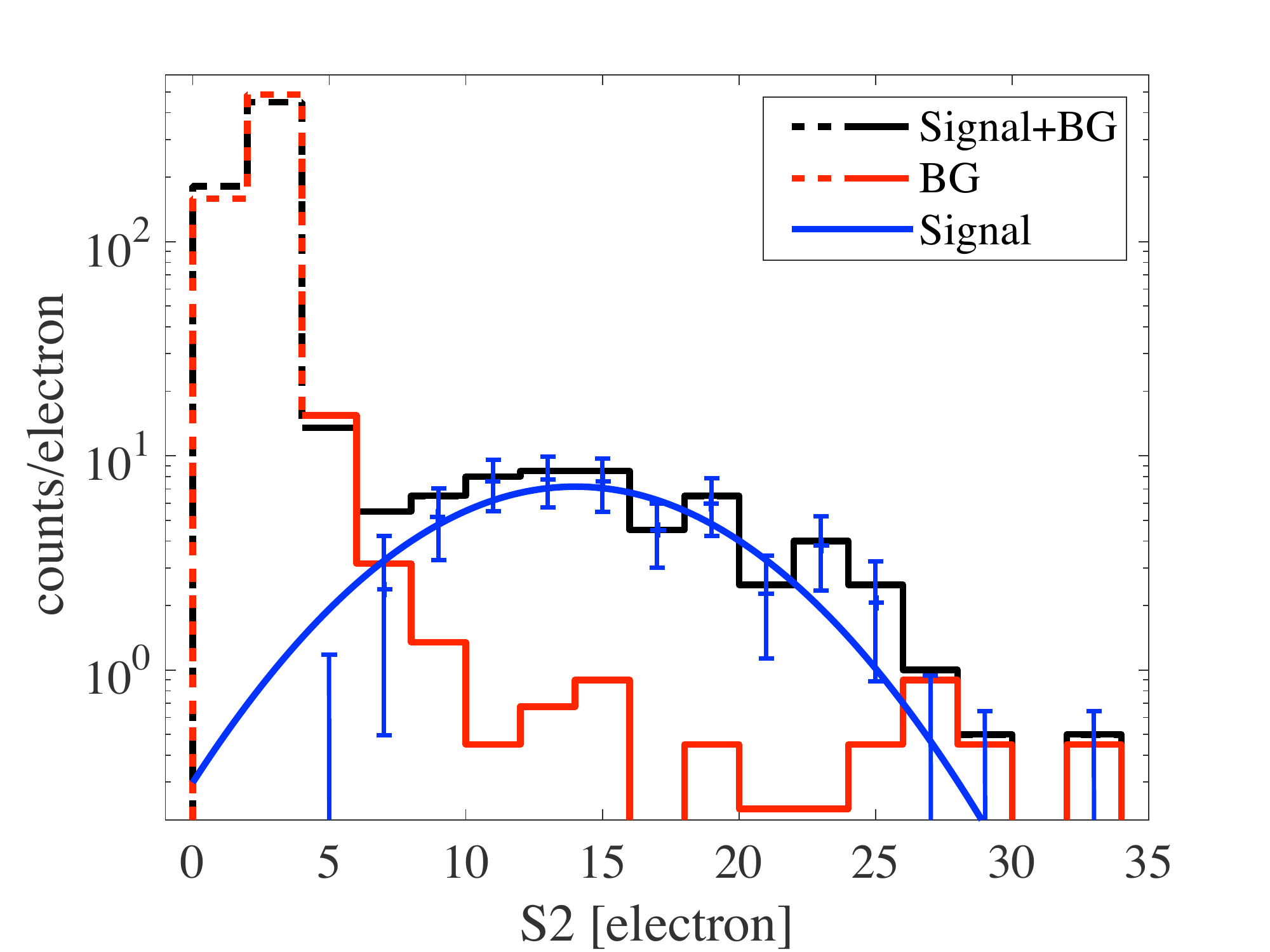}
\vskip -0.1cm
\caption{The black (solid and dashed) histogram shows the N-shell X-ray charge signals (with background). The red (solid and dashed) lines are the data-driven background model. Both dashed black and red lines mainly populated by SEs are not used for signal extrapolation. The blue data points are the N-shell X-ray charge spectrum after background subtraction fit with a Gaussian function (blue curve).}
\vskip -0.5cm
\label{fig:frc_f4}
\end{center}
\end{figure}

The X-rays' ER charge spectra are shown in Fig.~\ref{fig:frc_f5}. 
The charge peaks, from right to left, are from K-, L-, M-, and N-shell X-rays, respectively. 
The peaks are isolated by selecting events both to the upper-left of the line S$2_1=~$S$2_2$ in Fig.~\ref{fig:frc_f1} and with a second vertex S2 size within $\pm$2$\sigma$ of the 203~keV gamma-ray band mean.
% *** DQ 170323 - comment out the following section
%The N-shell peak is extracted with a different technique described in the next sub-section (Sec.~\ref{sec:n}). The shoulder to the right of the K-shell peak includes the bent tail events on the scatter plot, as discussed earlier. The continuum between the K- and L-shell peaks can be explained with the same ``energy transfer'' process that occurs for the K-shell events. 
The N-shell X-ray charge spectrum is also shown alone in Fig.~\ref{fig:frc_f4} along with its measured background, the details of which will be discussed in Sec.~\ref{sec:n}.
Both means and widths are extracted by fitting the peaks with Gaussian functions and are tabulated in Table~\ref{tab:qy} together with predicted values from the NEST model (NEST v0.98)~\cite{MatthewNEST}. Fig.~\ref{fig:Qy} includes $\textit{Q$_y$}$ measurements made in this analysis along with those made in an analysis of the LUX tritium calibration~\cite{LUXTritium}.

%\subsection{Estimation of Backgrounds for N-Shell X-ray Measurement}\label{sec:n}
\subsection{Optimization of N-Shell X-ray Analysis}\label{sec:n}
The background for the N-shell X-ray analysis is dominated by single extracted electrons primarily associated with photoionization of impurities in LXe by photons from S1 signals~\cite{EDWARDS200854}. These background events in the data appear as SEs preceding a 203~keV gamma-ray S2, and are shown in Fig.~\ref{fig:frc_f1} as the lower half of the leftmost population labeled ``SE''.
This feature is not centered at log$_{10}$(S2$_1$~[electron])~=~0 on the $x$-axis because efficiency corrections (both free electron lifetime and electron extraction efficiency) applied to all S2s are also applied to SEs for consistency to put all signals on the basis of the equivalent number of electrons at the initial interaction site. To resolve the N-shell X-ray peak shape with interference from background populations subtracted, a data-driven background model is established. 

Additionally, the population of signal plus background events for this N-shell X-ray analysis is selected using the 203~keV gamma-ray total reconstructed energy using S1 and S2 instead of its S2 size only as in the previous section.
Because S1 signal and S2 signal are anti-correlated as demonstrated in the following energy reconstruction model, the underlying electron-ion recombination fluctuation effect is canceled in the reconstructed energy by including the S1 signal. This results in a significantly better energy resolution than that achieved with S2-only energy scale.
This improved resolution helps suppress the background event population and increase the detector sensitivity to N-shell X-ray signals via a more effective selection of events, details of which are presented below. This additional technique is not required for K-, L-, and M-shell X-ray analyses because their backgrounds are negligible as discussed earlier.
%To establish the background model for the N-shell X-ray measurement, events with a SE preceding a combined K-shell X-ray and gamma signal are selected via their reconstructed energy.
%The photoionization SE background spectrum is normalized by the ratio of S1 light production within each energy window.

The energy reconstruction model is given by
\begin{align} \label{eq:energy}
	E_{total} = W \cdot (n_\gamma + n_e),
\end{align}
where $W$ is the energy required to produce a scintillation photon (exciton) or an ionization electron (electron-ion pair) and has a value of ${13.7\pm0.2}$~eV~\cite{Dahl}; $n_\gamma$ and $n_e$ are the number of photons and electrons produced at the interaction site, respectively. S1 and S2 both have units of phd and are proportional to the number of photons and the number of electrons, respectively: 
\begin{align} \label{eq:s1}
	n_\gamma = \frac{S1}{g_1},
\end{align}
\begin{align} \label{eq:s2}
	n_e = \frac{S2}{g_2}.
\end{align}
Parameters $g_1$ and $g_2$ are signal gains for S1 and S2 with respective values of $0.117\pm0.003$ phd/photon and $12.06\pm0.84$ phd/electron during the WS2013 period~\cite{LUXprd}. Parameter $g_2$ is the product of electron extraction efficiency (0.49$\pm$0.03) at the liquid-gas interface and the mean response to the single extracted electron ($24.66\pm0.02$ phd).
%The optimal S2 threshold for N-shell analysis is found to be 250~phd, which is the same as the K-, L-, and M-shell analysis S2 threshold. 

\begin{figure}[htbp]
\begin{center}
\vskip -0.1cm % -0.5
\includegraphics[width=0.45\textwidth]{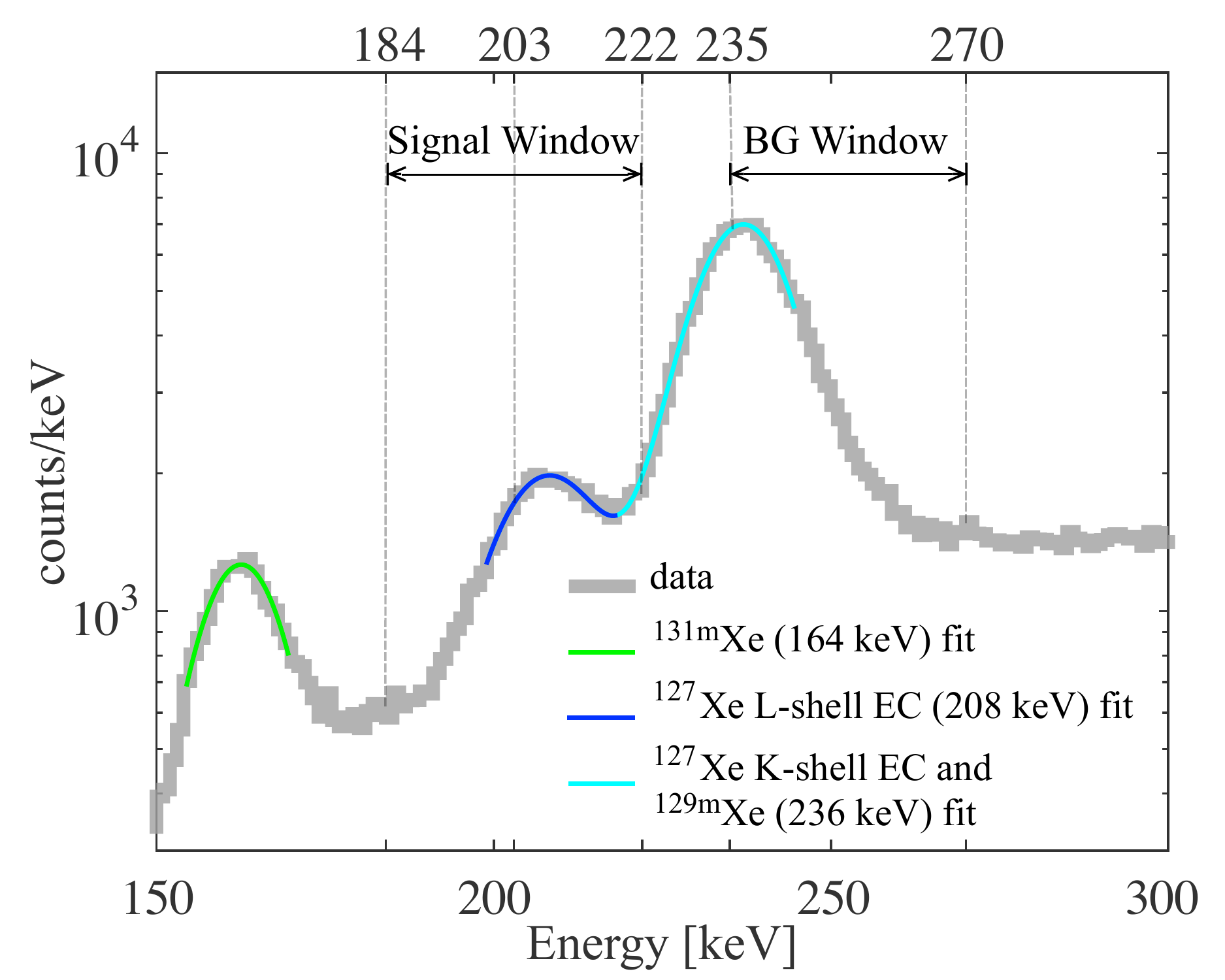}
\vskip -0.1cm
\caption{The reconstructed energy spectrum of single-scatter events in the energy range of interest. The ``Signal Window'' indicates the energy range where N-shell X-ray signals (Fig.~\ref{fig:frc_f4} solid blue) are exploited, while ``BG Window'' indicates the energy range where the N-shell X-ray calibration background model (Fig.~\ref{fig:frc_f4} solid red) is defined.}
\vskip -0.5cm
\label{fig:frc_f3}
\end{center}
\end{figure}

The reconstructed energy spectrum for
%single scatter 
events in the energy region of interest with a gamma-ray S2 threshold set at 250~phd is shown in Fig.~\ref{fig:frc_f3}. The first peak from the left consists of $^{131\textrm{m}}$Xe decay events with an energy of 164~keV mainly via IC. Events in the second peak centered at 208~keV are mainly L-shell EC events with a combined L-shell X-ray and gamma-ray S2. The third peak centered at 236~keV includes K-shell events with a combined K-shell X-ray and gamma-ray S2, as well as $^{129\textrm{m}}$Xe decay events with an energy of 236.1~keV mainly via IC. The peak mean values attained by Gaussian fitting agree well with the respective expected values of 208.1~keV and 236.1~keV to within 0.2{\%}. An analysis of the yields from these composite X-ray and gamma-ray events is presented in~\cite{EvanYiled}.

Due to energy resolution and low statistics, M- and N-shell EC event peaks are not visible in Fig.~\ref{fig:frc_f3}. For N-shell EC events, because the X-ray energy is negligible compared to that of the gamma ray, the underlying distribution is expected to be centered at 203~keV with a width similar to both the K-shell and L-shell peaks, given they are close in energy.
%The Signal+BG spectrum (blue stair) shown in Fig.~\ref{fig:frc_f4} includes all sub-threshold S2 pulses that occurs ahead of S2 in drift direction in all selected single scatter events. The events are selected with their energy within 2 sigma of the underlying energy distribution mean of N-shell EC events on both sides (Signal Window in Fig.~\ref{fig:frc_f3}). The BG spectrum (red stair in Fig.~\ref{fig:frc_f4}) is with energy in the region that is more than 3 sigmas to the right of the distribution mean of the N-shell EC events (BG Window in Fig.~\ref{fig:frc_f3}). 

As shown in Fig.~\ref{fig:frc_f3}, events for the N-shell $Q_y$ measurement are selected from within $\pm$2$\sigma$ of the 203 keV peak in combined energy (``Signal Window''). Small S2 pulses of potential N-shell X-ray signals preceding gamma-ray S2s are exploited from these events. A background spectrum of pulses preceding combined X-ray and gamma-ray S2s is made using events in the energy region more than 3$\sigma$ above the 203 keV peak (``BG Window''), which should be free of any N-shell EC events while still giving rise to the same photoionization SE background discussed earlier. With these selection criteria applied, Fig.~\ref{fig:frc_f4} shows the observed charge spectrum for X-ray S2 pulses. In this figure, Signal+BG events (black histogram) and the BG spectrum (red histogram) are drawn from the respective Signal and BG Windows shown in Fig.~\ref{fig:frc_f3}.
The BG spectrum is normalized by the ratio of the number of events in both selected energy ranges as well as the photoionization-based SE rate for different energies.
A distinct peak (solid black in Fig.~\ref{fig:frc_f4}) in the Signal+BG spectrum containing the majority of the N-shell X-ray charge signals is observed following the SE population (dashed black) with $>$5$\sigma$ significance above background.
The SE population present in the Signal+BG spectrum is well-modeled by the corresponding SE population in the BG spectrum (dashed~red).

%Corrections due to finite electron lifetimes arising from LXe impurities are applied to all pulses, including SE pulses, for consistency.
%These corrections result in a shift in the observed SE peak mean upwards from the true S2 size of a single extracted electron.
The N-shell X-ray Signal spectrum is obtained by subtracting the BG spectrum from the Signal+BG spectrum, and the Signal mean and width (Table~\ref{tab:qy}) are extracted by fitting the spectrum with a Gaussian function via the chi-squared method. 

A dedicated event visual assessment of 300 indicates that the LUX analysis code has a $\geq 99\%$ flat efficiency for N-shell X-ray signal down to a single electron with $90\%$ confidence level.

The best fit gives ${\chi}^2 = 6.7$ with 8 degrees of freedom. The probability of ${\chi}^2 \geq 6.7$ is $57\%$, which is reasonable. The best fit mean implies an average of %
$14\pm1$~electrons
produced at the ER interaction site for a given 186~eV energy recoil.

\subsection{Estimation of Systematics in Peak Shapes}\label{sec:systematics}
In Fig.~\ref{fig:frc_f1}, the bent tail to the right of the K-shell EC events on the 203~keV band is likely caused by an ``energy deposition transfer'' between the X-ray and gamma-ray signals within a small fraction of K-shell events. 
The 203~keV gamma ray deposits a small portion of its energy near the EC site via Compton scattering before later being fully absorbed. This leaves the Compton signal merged with the X-ray signal. 
A similar process that also contributes to the tail is the decay of the 203~keV state via two transitions with 145.4~keV gamma-ray and 57.6~keV gamma-ray (or IC electron) emissions (Fig.~\ref{fig:decay_scheme_127Xe}), of which the 57.6~keV signal is combined with the X-ray signal due to a shorter MFP.

The population that lies on the crossing of the line 
S$2_1=~$S$2_2$ 
and 375~keV band is populated by events that de-excite via two gamma-ray emissions (or one gamma-ray and one IC electron) of energy 172~keV and 203~keV from the 375~keV state (Fig.~\ref{fig:decay_scheme_127Xe}), with one of the two signals merged with the following X-ray signal. The symmetry feature is again due to the isotropic distribution of gamma-ray emission direction. The prominent population at the bottom right corner of both 203~keV and 375~keV bands consists of events with a combined X-ray and gamma-ray S2 followed by an isolated tail pulse that are incorrectly classified as a separate S2. 

The near-circular shape of the K-shell EC event distribution with log-log scale in Fig.~\ref{fig:frc_f1} suggests that the detector has comparable charge resolution between 33.2~keV (X-ray) and 203~keV (gamma-ray) energy depositions. At 203~keV, the recombination probability is smaller~\cite{EvanYiled}, but the recombination fluctuation, a significant component to the S2 energy resolution in LXe, is larger~\cite{EvanYiled}.

In Fig.~\ref{fig:frc_f5}, the shoulder to the right of the K-shell peak includes the bent tail events on the scatter plot, as discussed earlier. The continuum between the K- and L-shell peaks can be explained with the same ``energy deposition transfer'' process that occurs for the K-shell events.

\begin{table}[h]
\caption{$\textit{Q$_y$}$ mean and width comparisons between data and NEST (v0.98) prediction for each X-ray. The first error quoted in measurement is the statistical error and the second is the systematic error.}
\begin{tabular}{l|r@{~~~}l@{~$\pm$~}l@{~$\pm$~}l@{~~}r@{.}l@{~~}l@{~~}l}
\hline \hline
& Energy  & \multicolumn{3}{l}{$\textit{Q$_y$}$ [$n_e$/keV]} & \multicolumn{4}{l}{Width($\sigma$)} \\ \hline
Data & 33.2 keV   &   22.72 & 0.03 & 1.58  &   3&62 &$\pm$~0.02&$\pm$~0.25 \\
     &  5.2 keV   &    30.8 & 0.1 & 2.1    &   6&28 &$\pm$~0.09 & $\pm$~0.44 \\
     &  1.1 keV   &    61.4 & 0.5 & 4.3    &   12&9 &$\pm$~0.4  & $\pm$~0.9 \\
     &  186 ~eV   &    75.3 & 6.5 & 5.2    & \multicolumn{1}{r}{30}& & $^{+~4}_{-~3}$ &$\pm$~ 2 \\
\hline
%NEST & 33.2 keV & \multicolumn{3}{l}{23.1} &   \multicolumn{4}{l}{~3.4} \\
NEST & 33.2 keV & \multicolumn{3}{l}{23.1} &   3&\multicolumn{3}{l}{4} \\
     & 5.2 keV & \multicolumn{3}{l}{33.2} &   5&\multicolumn{3}{l}{2} \\
     & 1.1 keV & \multicolumn{3}{l}{54.5} &   12&\multicolumn{3}{l}{3} \\
     & 186 ~eV & \multicolumn{3}{l}{65.4} &   32&\multicolumn{3}{l}{5} \\
%& 5.2 keV & 33.2 & 5.2 \\
%& 1.1 keV & 54.5 & 12.3 \\
%& 186 ~eV & 65.4 & 32.5 \\

\hline \hline
\end{tabular}
\label{tab:qy}
\end{table} 

\begin{figure}[htbp]
\begin{center}
\vskip 0.1cm % -0.5
\includegraphics[width=0.45\textwidth]{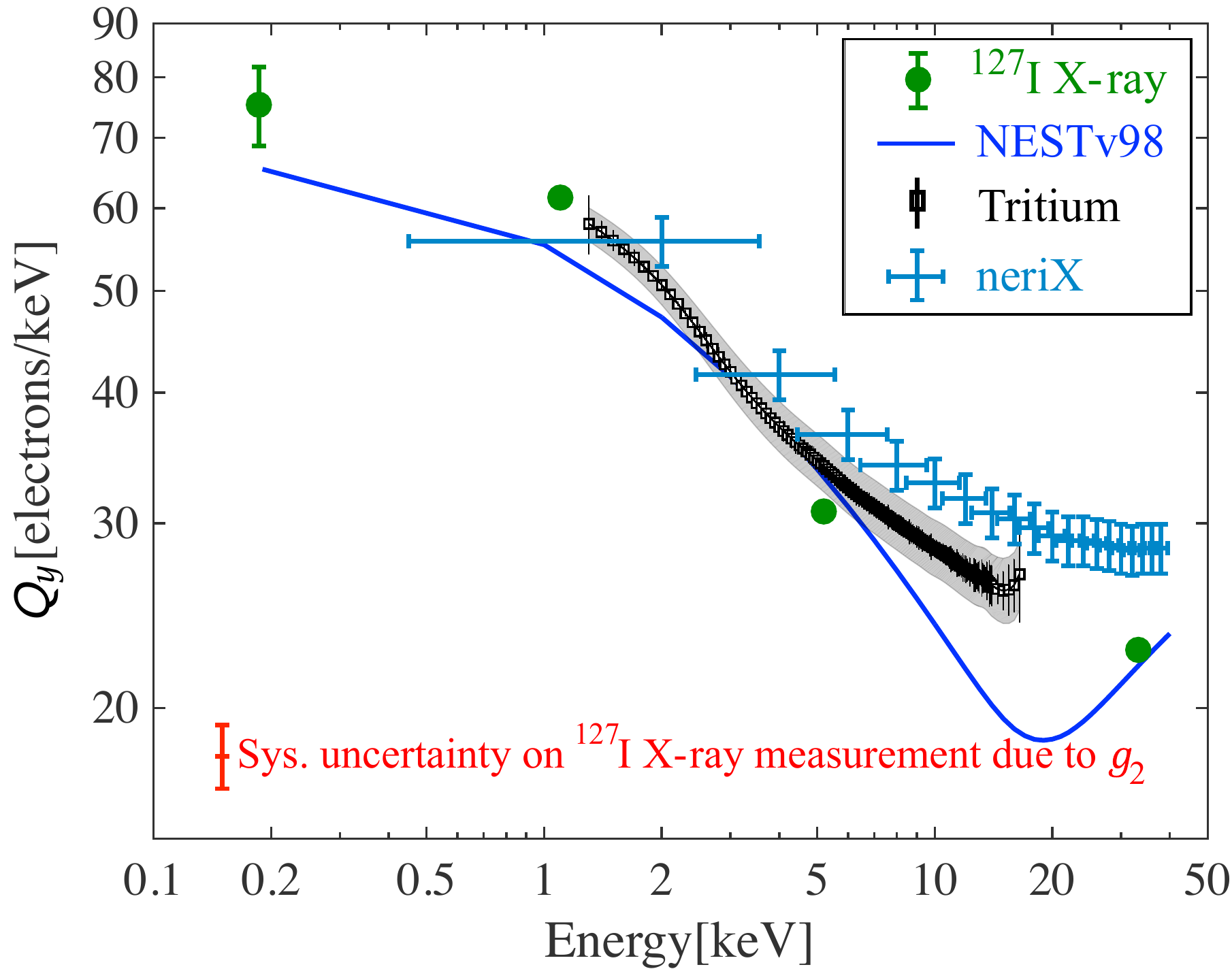}
\vskip -0.1cm
\caption{The $\textit{Q$_y$}$ mean values (green points) measured from $^{127}$Xe EC events (this work) at 180 V/cm. The red bar indicates the systematic uncertainty on the $^{127}$Xe measurement due to $g_2$. Also shown are the $\textit{Q$_y$}$ measured from the LUX tritium calibration~\cite{LUXTritium} at 180 V/cm, of the NEST v0.98 model~\cite{MatthewNEST} at 180 V/cm, and measured by the neriX experiment~\cite{Goetzke:2016lfg} at 190 V/cm. The light gray band on the tritium measurement indicates the systematic uncertainty due to $g_2$. %and the light blue band on the NEST indicates systematic uncertainty in the model.
% *** DQ 161108 RG We should only show a red bar, not a circle - what does the circle mean?
}
\vskip -0.5cm
\label{fig:Qy}
\end{center}
\end{figure}

\section{discussion and conclusion}\label{sec:conclusion}
The $\textit{Q$_y$}$ mean and width measured from each X-ray are listed in Table~\ref{tab:qy} along with quoted statistical and systematic uncertainties. The systematics are dominated by the uncertainty in the electron extraction efficiency at the liquid-gas interface. Because the efficiency has the same effect on all charge signals, the systematics shift all four measured means in the same direction simultaneously. 

The table also lists the corresponding NEST v0.98 predictions for comparison. There is good agreement between the measured K- and L-shell $\textit{Q$_y$}$ means and NEST v0.98 predictions. For the M- and N-shell, while they still agree within uncertainty of $2\sigma$ level by taking systematics into account, the measured means' central values are found to be 12.7{\%} and 15.9\% higher than NEST v0.98 predictions.

As shown in Fig.~\ref{fig:Qy}, there is slight tension between L-shell measured mean and LUX tritium measurement~\cite{LUXTritium}, given similar systematic on both measurements due to $g_2$. This can be possibly understood as different track topologies for X-ray and beta particle in LXe. The M-shell measurement reasonably agree with inferred trend of tritium measurement, indicating smaller difference in track topologies for lower energies.

If we assume a constant $W$-value of 13.7~eV independent of energy, the N-shell measurement 
indicates that nearly all of the energy for 186~eV electron recoil appears as electrons. 
%
% Average 13.5 excitations per 186 eV event for w = 13.7 eV
% Charge yield 75 elec/keVee * 0.186 = 14.0 excitations
% 
With this assumption, 
we are able to place upper limits on both electron-ion recombination probability and photon emission probability for 186~eV electron recoils.
%indicates small production of excitons on the ER track and small electron-ion recombination for a 186~eV electron recoil. 
The N-shell measured charge yield indicates a $90\%$ confidence upper limit of 0.056 for electron-ion recombination probability, 
if we assume the theoretical estimate for $n_{ex}/n_{ion}$ for ER is 0.06 ~\cite{alphavalue1,alphavalue2}.
The constraint from our observation on the electron-ion recombination probability becomes even more severe if we assume $n_{ex}/n_{ion}$ for ER is the 0.2 measured for higher energy sources (662~keV in~\cite{alphMeasure2} and 1~MeV in~\cite{Doke2002}).
%For reference, the existing estimates of $n_{ex}/n_{ion}$ for ER are 0.06 by theory~\cite{alphavalue1,alphavalue2} and 0.2 measured with higher energy sources (662~keV in~\cite{alphMeasure2} and 1~MeV in~\cite{Doke2002}).
Our observed N-shell charge yield also places a $90\%$ confidence upper limit for the photon emission probability per event at 186~eV of 0.11. 

% The zero recombination for an N-shell event can be understood to result from so few electrons being present on the ionization track that the recombination probability with other ions is negligible.

% The L- and M-shell measured means also reasonably agree with LUX tritium $\textit{Q$_y$}$ measurements~\cite{LUXTritium} and its inferred trend (Fig.~\ref{fig:Qy}). 

For line widths, the table shows good agreement between measurements and NEST v0.98 predictions. The NEST width predictions are calculated via NEST with simulation of LUX detector effects such as binomial electron extraction, light collection efficiencies, and SE size resolution~\cite{LUXprd} during the WS2013 data taking period. The widths are mainly comprised of two components, the intrinsic electron-ion recombination, and the following binomial processes due to electron drift lifetime and the electron extraction efficiency. For N- and M-shell, the widths are dominated by the binomial processes because of the relatively small number of electrons produced at the ER sites. 
The measured $\textit{Q$_y$}$ indicates small recombination probability as discussed above. 
For L-shell events, both the intrinsic recombination and the following binomial processes become significant effects in determining the width. For K-shell events, the intrinsic recombination becomes the dominant contributor to the width due to the significance of recombination fluctuations at higher energy.

%However, for low and ultra-low energies such as 1.1 keV and 186 eV, the electron-ion recombination no longer contributes to their widths as the recombination probability goes down to near zero as discussed. What drives the widths at this point apart from extraction efficiency is the SE resolution, 24.1{\%}~\cite{LUXprd} measured in LUX Run3. Given only tens of electrons produced at interaction site, the per-electron based size variation contributes to the overall width of charge distribution. The SE resolution is irrelevant at higher energies, because the upper fluctuating electrons are canceled by downward fluctuating ones on average with hundreds of electrons.

In conclusion, we have successfully extracted K- (33.2 keV), L- (5.2 keV), M- (1.1 keV), and N- (186 eV) shell X-ray electronic recoil (ER) charge signals due to $^{127}$Xe electron capture decay in the LUX detector LXe active volume from the WS2013 dataset. Both the mean and sigma of $\textit{Q$_y$}$ associated with each energy are accurately measured. The N-shell X-ray $\textit{Q$_y$}$ measurement with 186 eV electronic recoil energy deposition represents the lowest-energy ER $\textit{in situ}$ measurements that have been explored in Xe. 

%The calibration results concluded in this work are used to tune the existing NEST v0.98 model, which is then implemented to model the ER response for the improved LUX WS2013 WIMP search analysis~\cite{LUXRun3Reanalysisprl}.

\section{acknowledgements} \label{sec:acknowledgements}
% ***DQ 161108 RG Where did this come from?? How recent is it? - can you include a comment in TeX that makes it clearer?
% ***RG 161115 DQ this is taken from Run3 reanalysis paper
This work was partially supported by the U.S. Department of Energy (DOE) under Awards No. DE-FG02-08ER41549, No. DE-FG02-91ER40688, No. DE-FG02-95ER40917, No. DE-FG02-91ER40674, No. DE-NA0000979, No. DE-FG02-11ER41738, No. DESC0006605, No. DE-AC02-05CH11231, No. DE-AC52-07NA27344, No. DE-FG01-91ER40618, and No. DE-SC0010010; the U.S. National Science Foundation under Grants No. PHYS-0750671, No. PHY-0801536, No. PHY-1004661, No. PHY-1102470, No. PHY-1003660, No. PHY-1312561, No. PHY-1347449, No. PHY-1505868, No. PHY-1636738, and No. PHY-0919261; the Research Corporation Grant No. RA0350; the Center for Ultra-low Background Experiments in the Dakotas (CUBED); and the South Dakota School of Mines and Technology (SDSMT). LIP-Coimbra acknowledges funding from Funda\c{c}\~{a}o para a Ci\^{e}ncia e Tecnologia (FCT) through the Project-Grant No. PTDC/FIS-NUC/1525/2014. Imperial College and Brown University thank the UK Royal Society for travel funds under the International Exchange Scheme (IE120804). The UK groups acknowledge institutional support from Imperial College London, University College London and Edinburgh University, and from the Science \& Technology Facilities Council for PhD studentships No. ST/K502042/1 (AB), No. ST/K502406/1 (SS), and No. ST/M503538/1 (KY). The University of Edinburgh is a charitable body, registered in Scotland, with Registration No. SC005336.

This research was conducted using computational resources and services at the Center for Computation and Visualization, Brown University. 

We acknowledge the work of the following engineers who played important roles during the design, construction, commissioning, and operation phases of LUX: S. Dardin from Berkeley, B. Holbrook, R. Gerhard, and J. Thomson from UC Davis, and G. Mok, J. Bauer, and D. Carr from Livermore.

We gratefully acknowledge the logistical and technical support and the access to laboratory infrastructure provided to us by the Sanford Underground Research Facility (SURF) and its personnel at Lead, South Dakota. SURF was developed by the South Dakota Science and Technology Authority, with an important philanthropic donation from T. Denny Sanford, and is operated by Lawrence Berkeley National Laboratory for the Department of Energy, Office of High Energy Physics.

\bibliographystyle{apsrev4-1}

\bibliography{run3_127xe_cali}

\end{document}